\pdfoutput=1 
\documentclass[12pt]{article}
\usepackage{amsmath,amsthm,amssymb,enumerate}
\usepackage[authoryear]{natbib}
\usepackage{booktabs,multirow}
\usepackage[left=1in,top=1in,right=1in,bottom=1in]{geometry} 
\usepackage{natbib}
\usepackage{setspace}
\usepackage{dcolumn,comment}
\usepackage{pbox}
\usepackage{caption,subfig}
\usepackage{hyperref}
\usepackage[title]{appendix}

\newcolumntype{d}[1]{D{.}{.}{-1} }

\usepackage{tikz}
\usetikzlibrary{shapes.geometric, arrows}
\tikzstyle{process2} = [rectangle, minimum width=1cm, minimum height=1cm, text centered]
\tikzstyle{process} = [rectangle, minimum width=1cm, minimum height=1cm, text centered, draw=black]
\tikzstyle{processc} = [rectangle, minimum width=1cm, minimum height=1cm, text centered, draw=black, fill=gray!30]
\tikzstyle{arrow} = [thick,->,>=stealth]

\def\b0{{\bf 0}}

\def\bSig\mathbf{\Sigma}

\doublespace

\def\cw0{c_{w_0}}

\singlespace
\title{Evaluating Effectiveness of Public Health Intervention Strategies for Mitigating COVID-19 Pandemic}
\author{\small Shanghong Xie$^1$, Wenbo Wang$^2$, Qinxia Wang$^1$, Yuanjia Wang$^{1}$\thanks{Yuanjia Wang, Department of Biostatistics, Mailman School of Public Health, Columbia University, New York, NY, U.S.A. Email: yw2016@cumc.columbia.edu}, Donglin Zeng $^{2}$\thanks{Donglin Zeng, Department of Biostatistics, University of North Carolina at
Chapel Hill, Chapel Hill, North Carolina, U.S.A.  Email: dzeng@email.unc.edu}}

\date{\footnotesize $1.$ Department of Biostatistics, Columbia University, New York, NY, U.S.A. \\
$2.$ Department of Biostatistics, University of North Carolina at
Chapel Hill, Chapel Hill, North Carolina, U.S.A.}

\begin{document}
\maketitle

\begin{abstract}
Coronavirus disease 2019 (COVID-19) pandemic is an unprecedented global public health challenge. In the United States (US), state governments have implemented various non-pharmaceutical interventions (NPIs), such as physical distance closure (lockdown), stay-at-home order, mandatory facial mask in public in response to the rapid spread of COVID-19. To evaluate the effectiveness of these NPIs, we propose a nested case-control design with propensity score weighting under the quasi-experiment framework to estimate the average intervention effect on disease transmission across states. We further develop a method to test for factors that moderate intervention effect to assist precision public health intervention. Our method takes account of the underlying dynamics of disease transmission and balance state-level pre-intervention characteristics.  We prove that our estimator provides causal intervention effect under assumptions.  We apply this method to analyze US COVID-19 incidence cases to estimate the effects of six interventions. We show that lockdown has the largest effect on reducing transmission and {reopening bars} significantly increase transmission. States with a higher percentage of non-white population are at greater risk of increased $R_t$ associated with reopening bars.
\end{abstract}


\doublespace

\section{Introduction}
\label{s:intro}
Coronavirus disease 2019 (COVID-19) pandemic is an unprecedented global health crisis that has brought tremendous challenges to humanity. 
Countries around the world have introduced mitigation measures and non-pharmaceutical interventions (NPIs) to respond to the crisis before vaccines are widely available. Within the United States (US), there is tremendous heterogeneity in terms of when mitigation strategies were implemented and lifted across states  and a varying-degree of combinations of containment, social distancing,  and lockdown (i.e., physical distance closures including closure of schools and businesses). Decisions for implementing these strategies partially rely on essential statistics and  epidemiological models that characterize the course of COVID-19 outbreak. However, despite numerous disease forecast models proposed in literature \citep{Ray2020medrxiv}, there is a   lack of methods to evaluate intervention effects that are {robust and generalizable} to accurately account for the heterogeneity between populations. There is no study on precision NPIs that are tailored to states and high risk populations susceptible to COVID-19 infection. Thus, it is imminent to  study average treatment effect and heterogeneity of treatment effect (HTE) to inform health policy on COVID-19 responses. 

One essential component of evaluating an NPI is to identify a proper outcome measure. During COVID-19 pandemic, daily cases and deaths are reported in each state in the US. However, it is well known that there are  a large number of pre-symptomatic  cases accounting for about 40\% of transmissions \citep[CDC;][]{oran2020prevalence} and there has been a shortage of accurate {polymerase chain reaction (PCR)} tests especially during the early phase of the pandemic. In addition to  lagged reports, the observed cases do not fully reflect how the epidemic evolves in real time, so simply using reported cases or deaths as outcomes may lead to suboptimal decisions. In contrast, mechanistic-based epidemiological models can  estimate the true underlying  dynamics of COVID-19 epidemic and provide the time-varying effective reproduction number ($R_t$) as a outcome measure. In particular, our earlier work \citep{wang2020survival} proposed to combine nonparametric statistical curve fitting with infectious disease epidemiological models of the  transmission dynamics. This model accounts for pre-symptomatic transmission of COVID-19 and provides estimates of infection rates and reproduction numbers. These quantities, when modeled as time-varying, can effectively capture the underlying dynamics that  govern the disease transmission, and are the appropriate measures that should be targeted by an intervention. For example, a reproduction number below one indicates that the disease epidemic is under control. Thus, we use time-varying  reproduction number, denoted by $R_t$, as the outcome measure of the intervention effect.

To estimate intervention effects on COVID-19, we consider methods that use natural experiment designs to allow drawing causal inference under assumptions. Since different states implemented interventions at different time points,  the effects of mitigation strategies can be treated as quasi-experiments where subjects receive distinct interventions before or after the initiation of the intervention.  The longitudinal pre-post intervention designs including regression-discontinuity design \citep{hahn2001identification} and difference in difference (DID) regressions are frequently used in practice to analyze quasi-experiments data \citep{wing2018designing, leatherdale2019natural}. Regression-discontinuity design  estimates intervention effects at the local point at which the probability of implementing the intervention changes discontinuously. DID estimates the intervention effect by examining the interaction term between time and group in a regression model. It allows valid inference assuming that outcome trends are parallel in treated and untreated group and local randomization holds (i.e., whether a subject falls immediately before or after the initiation date  of an intervention may be considered random, and thus the ``intervention assignment" may be considered to be random). When the first assumption  does not hold, synthetic control \citep{abadie2010synthetic} is proposed to weight observations so that pre-intervention average effects are similar between groups. 

Several recent works have investigated the intervention effects of COVID-19 mitigation strategies. Process-based infectious disease models  are used to simulate counterfactual outcomes under different manipulations of model parameters and assumptions on the intervention effects \citep{ferguson2020impact,Pei2020Differential}. These models may be useful to simulate disease outcomes under hypothetical scenarios of interventions, but do not estimate intervention effects based on observed data. \cite{auger2020association} and \cite{rader2021mask} evaluated the associations between the interventions and outcomes (i.e., cases, deaths, and $R_t$) by regression models.  \cite{davies2020effects} and  \cite{flaxman2020estimating} assessed the intervention effects by modelling the basic reproduction number $R_0$ or $R_t$ as intervention dependent. These approaches included state-level characteristics as covariates in the model, but did not  investigate the causal effects. \cite{cho2020quantifying} considered synthetic  control  and  DID approach by fitting linear regression with reported cases and deaths as outcomes, but did not take account of the  dynamic feature of the disease transmission.

In this paper, we propose a novel method to assess the effect of NPIs using the reported daily cases from each state in US. Compared to existing literature, our work has several new aspects as follows.
First,  since COVID-19 outbreak may occur at different times in each state, calendar time may not be a good  measure of the stage of epidemic. To create a meaningful time horizon that reflects each state's epidemic course when comparing intervention effects, we  align states by transforming calendar time to time since the first reported case.  Second, we  use a nested case-control design (e.g., treating the implementation of an intervention as an event) \citep{ernster1994nest} and propensity score weighting to estimate intervention effect. Specifically, for each state that has implemented an intervention at a given time point, we define a set of control states as those which have not yet implemented the intervention. Therefore, a state that  implements a policy at a later time  can serve as control for other states that have acted earlier. This design would allow observations from  different time periods in the same state to serve in  both treated and untreated groups, so that the longitudinal data from 50 states can be efficiently used. Third, to balance treated and untreated groups, we construct propensity scores using pre-intervention characteristics including state-level social demographic variables (e.g., social vulnerability index (SVI), state's average age and race distribution) as well as time-varying characteristics of  the epidemic (e.g., pre-intervention case rate, hospitalization, $R_t$). We prove that our estimator yields the causal effect of an intervention under assumptions. Lastly, we further estimate heterogeneity of treatment effect (HTE) using estimation equations that include important hypothesized moderators. The developed method is applied to analyze US COVID-19 data  to estimate the effects of six NPIs. We show that the lockdown during spring of 2020 had the largest effect on reducing $R_t$ and {reopening bars} led to significant increase of  disease transmission.

\section{Method for Evaluating Intervention Strategies}

\subsection{Outcome measure for estimating NPI effects}\label{sec:outcome}

To estimate the time-sensitive infection rate or reproduction number as an outcome for assessing NPIs, we adopt a previously developed method, survival-convolution model {\citep{wang2020survival}},  over days since the reported first case. This model is inspired by the epidemiological Susceptible-Exposed-Infective-Recovered (SEIR)  model, but has fewer assumptions and model parameters, and demonstrate adequate prediction performance among an ensemble of models in the CDC forecast task (\url{https://www.cdc.gov/coronavirus/2019-ncov/covid-data/forecasting-us.html}).

To be  specific, let $N_{i}(t)$ be the number of individuals in the $i$th state who are newly infected by COVID-19 at time $t$. Since we are  interested in the overall population-level disease transmission epidemiology, we  assume that the virus transmits from one individual to another at the same rate at a given time $t$.  In this population, the duration of an individual remaining infectious in the epidemic is from a homogeneous distribution at any calendar time  $t$ (in days).  Let $t_j$ denote the time when individual  $j$ in this population is infected ($t_j=\infty$ if never infected), and let $G_{j}$ be the duration of this individual remaining infectious to any other person and staying in the transmission chain. 
Since the total number of individuals who are newly infected at time $(t-m)$ is $N_{i}(t-m)$, the number of  individuals who remain infectious $m$ days later (i.e., at time $t$) is
$N_{i}(t-m) S(m),$
where $S(m)$ denotes the proportion of persons remaining infectious after $m$ days of being infected, or equivalently,  the survival probability at day $m$ for $G_j$. Thus, at  time $t$, the total number of individuals who can infect others is  $$M_i(t)=\sum_{m=0}^{\infty} N_{i}(t-m) S(m).$$
On the other hand, right after day $t$, some individuals will no longer be in the transmission chain {due to testing positive and quarantine or out of infectious period} (i.e., duration $G_j=(t-t_j)$), and the total number of these individuals denoted by $W_i(t)$ is
$$W_i(t)=\sum_{j: j\textrm{ in  state $i$}} I(t_j\le t, G_j=t-t_j)=\sum_{m=0}^{\infty} \sum_{\textrm{$j$: $j$ is infected at $(t-m)$}}  I(G_j=m),$$ or equivalently
$$W_i(t)=\sum_{m=0}^{\infty}N_{i}(t-m)[S(m)-S(m+1)].\eqno(1)$$
Therefore, $M_i(t)-W_i(t)$ is the number of individuals who can still infect others after day $t$.
Assuming the infection rate at $t$ to be $a_i(t)$, then at day $(t+1)$ the number of newly infected patients is $$a_i(t)[M_i(t)-W_i(t)],$$ which yields
$$N_{i}(t+1)=a_i(t)\sum_{m=0}^{\infty} N_{i}(t-m)S(m+1). \eqno(2)$$

Equation (2)  gives a convolution update for the new daily cases using the past days' number of cases. {This equation considers  three important quantities to characterize COVID-19 transmission}: the initial date, $t_{0}$, of the first (likely undetected) case in the epidemic, the survival function of time to out of transmission, $S(m)$, and the infection rate over calendar time, $a_i(t)$.   {\cite{wang2020survival}} estimated $a_i(t)$ as a piece-wise linear function with knots placed at  intervention dates and every 2-3 weeks, and approximated the survival function $S(m)$ based on previous literature \citep{li2020early}. Similarly, {we computed $a_i(t)$ as piece-wise linear function, placing knots  at the state-specific intervention dates and every 2 weeks between interventions and modelled $S(m)$ as an exponential distribution.}  To estimate both $t_0$ and $a_i(t)$, 
\cite{wang2020survival} proposed to minimize a squared loss between the square-root transformed reported daily new cases and  predicted new cases from models (1) and (2).

Note that $a_i(t)$ is time-varying because the infection rate depends on how many close contacts one infected individual may have at day $t$, which is affected by NPIs (e.g., stay-at-home order, lockdown) and saturation level of the infection in the whole population. With the number of new infections $N_i(t)$ estimated from survival-convolution model in (2), the   effective  reproduction number {\citep{cori2013new}} is  defined as 
$$
R_{it}=\frac{N_i(t)}{\sum_{k=0}^{\infty}w_k N_i(t-k)},\eqno(3)$$
which is the number of secondary infections caused by a primary infected individual in a population at time $t$ while accounting for the entire incubation period of the primary case.  Thus, $R_{it}$  measures temporal changes of the disease transmission. Here, $w_k$ is the probability mass function of the distribution of serial intervals for SARS-CoV-2 ({a Gamma distribution}), which is obtained from  \cite{nishiura2020serial} and \cite{scire2020reproductive}.   

\subsection{Average intervention effect (ATE) and assumptions}

For the ease of presentation, we focus on a particular intervention (lockdown, for instance) in this section.
Our goal is to estimate the overall average effect of the intervention across states. To define the causal estimand, we introduce the following notations to  define a time-specific intervention effect. For any time period $\Delta>0$ such that the probability of two states implementing the intervention  within $\Delta$ days is zero, we let $Y_i^{(1)}(t+\Delta; t)$ denote the potential  change of the  reproduction number between $t$ and $(t+\Delta)$, had the intervention been applied at time $t$ and had there been no  other interventions between time $t$ and $t+\Delta$. Let $Y_i^{(0)}(t+\Delta;t)$ be the same potential outcome when there was no intervention at time $t$. Correspondingly, the time-specific intervention effect is defined as
$$\gamma(\Delta, t)=E[Y_i^{(1)}(t+\Delta;t)-Y_i^{(0)}(t+\Delta;t)].$$
In other words, we consider a hypothetical scenario where at time $t$, each state imposes the intervention and the other scenario where there is no such intervention at  $t$ and before. Then $\gamma(\Delta, t)$ is the difference between the change of the  reproduction number $\Delta$ days after time $t$. A negative value of $\gamma(\Delta,t)$ implies that the intervention at time $t$ can slow down the spread of the virus.
However, since very few states impose the intervention on the same day since disease outbreak, estimating $\gamma(\Delta,t)$ for each $t$ is not feasible. Instead, we define an average intervention effect (ATE) as the average of $\gamma(\Delta,t)$ over all possible intervention times, i.e., 
$$\gamma(\Delta)\equiv \int \gamma(\Delta,t)dF_T(t),$$
where $F_T(\cdot)$ is the distribution of the intervention time $T$.
Hence,  $\gamma(\Delta)$ can be viewed as an overall evaluation of the intervention effect over all states.
We are interested to estimate $\gamma(\Delta)$ using  empirical data.

For each state $i$, we set time zero to be its first reported case and let $Y_i(t+\Delta; t)$ be the change of  reproduction number  between $(t+\Delta)$ and $t$ {(i.e., $R_{it+\Delta}-R_{it}$)}, where the reproduction numbers are {estimated as Section \ref{sec:outcome}}. Let $X_i$ be the state-specific characteristics including a constant of one.  Let $T_i$ denote  the intervention time  and let $T_i=\infty$ if the state has never implemented this intervention. Let $F_T(t)$ denote the distribution of $T_i$, assumed to have a support on ${\cal T}$.  
To estimate $\gamma(\Delta)$ from observed data,
we require the following assumptions:
\\
(a) Suppose no other intervention occurs between $t$ and $t+\Delta$. We assume when $T_i=t$ (i.e., there is an intervention at $t$), $Y_i^{(1)}(t+\Delta;t)=Y_i(t+\Delta; t)$.
\\
(b) Suppose  no other intervention occurs between $t$ and $t+\Delta$ and the intervention of interest has not been imposed before $t$, we assume $Y_i^{(0)}(t+\Delta;t)=Y_i(t+\Delta; t).$
\\
(c) Assume no unobserved confounders: conditional on $T_i\ge t$, 
$T_i=t$ is independent of $Y_i^{(a)}(t+\Delta;t), a=0,1$ given $X_i$ and $H_i(t)$, where $H_i(t)$ denotes the observed epidemic history by time $t$.

Assumptions (a) and (b) are equivalent to the consistency assumption in causal inference. Both (a) and (b) also imply that there are no delayed effects from any other previous interventions prior to time $t$. This is plausible since the interventions do not occur frequently and the effects can decline rapidly, as seen by  multiple re-surges in this pandemic. Furthermore, even though the previous intervention may affect the infection rate at time $t$,  since the potential outcome of interest is the change of the infection rate or reproduction number since time $t$, the effect on this change can be much smaller.
Assumption (c) is the no-unobserved confounder assumption in causal inference literature. If all relevant epidemic history and other information associated with implementing an intervention at time $t$ are collected as $H_i(t)$ and $X_i$, this assumption holds. In our application, we will explore a list of candidate variables as ($X_i, H_i(t))$ and identify a subset data-adaptively.

Next, we justify why the assumptions enable us to estimate $\gamma(\Delta)$. First, under assumption (c),  we have 
$$\hspace{-2in}\gamma(\Delta, t)=
E\left[\frac{I(T_i=t)}{P(T_i=t|T_i\ge t, H_i(t), X_i)}\left\{Y_i^{(1)}(t+\Delta;t)\right\}\right]$$
$$ -E\left[\frac{I(T_i> t+\Delta)}{P(T_i> t+\Delta|T_i\ge t, H_i(t), X_i)}\left\{Y_i^{(0)}(t+\Delta;t)\right\}\right] .$$
Second, since  $P(T_i> t|T_i\ge t, H_i(t), X_i)=
P(T_i> t+\Delta|T_i\ge t, H_i(t), X_i)$ for any $t$ in the support of $F_T(t)$,  according to assumptions (a) and (b), the right-hand side is further equal to
$$\hspace{-2in}\gamma(\Delta, t)=E\left[\frac{I(T_i=t)}{P(T_i=t|T_i\ge t, H_i(t), X_i)}\left\{Y_i(t+\Delta;t)\right\}\right]$$
$$ -E\left[\frac{I(T_i> t+\Delta)}{P(T_i> t|T_i\ge t, H_i(t), X_i)}\left\{Y_i(t+\Delta;t)\right\}\right] .\eqno(4)$$
Therefore, if we posit a model for the intervention time $T_i$ given $H_i(t)$ and $X_i$, an inverse probability weighted estimator based on (4) can be used to estimate $\gamma(\Delta,t)$.  Equation (4) further provides a way to consistently estimate $\gamma(\Delta)$ by simply averaging the estimated $\gamma(\Delta,t)$ over all empirical intervention times from all states.

\subsection{Inference procedure for the average intervention effect}

The main idea for estimation is to create a separate set of control states for  ``case states" that implemented an intervention at a given time point and then aggregate over  case states. To balance pre-intervention differences between states, we will construct  propensity scores for case states that intervened at different time points, since eligible control states may differ. 
Specifically, in the first step, we estimate the propensity scores, $P(T_i=t|H_i(t), X_i)$ in (4), by fitting a logistic regression model, 
$$\textrm{logit}\left\{P(T_i=t|T_i\ge t, H_i(t), X_i)\right\}=(H_i(t), X_i)^T\beta,$$
where
$X_i$ contains all prognostic variables for the intervention at the baseline such as demographic distributions and SVI index, and $H_i(t)$ can be the average cases and deaths in the past week(s) before time $t$.
To estimate $\beta$, we solve the following estimating equation 
$$\sum_{i=1}^n \int  (H_i(t), X_i)^TI(T_i\ge t) \left[I(T_i=t)-\frac{\exp\{(H_i(t), X_i)^T\beta\}}{1+\exp\{(H_i(t), X_i)^T\beta\}}\right]d\widehat F_T(t)=0,$$
where
$\widehat F_T(t)$ denotes the empirical distribution of the intervention times. 
In detail,  if we use $X_{ij}$ to denote $(H_i(T_j), X_i)$ and $\delta_{ij}=I(T_i=T_j)$ , we can estimate $\beta$
by solving 
$$\sum_{i=1}^{n}\sum_{j\in S(i)}
X_{ij}
\left\{\delta_{ij}-\frac{\exp\{X_{ij}^T\beta\}}{1+\exp\{X_{ij}^T\beta\}}\right\}=0,$$
where $S(i)$ is a set of state $i$ and all other eligible control states (for example, states that have not implemented an intervention by $T_i$; similar to a nested case-control design when treating implementation of an intervention as the event).
Once we obtain the estimate for $\beta$, denoted by $\widehat\beta$, the propensity score for state $i$ at its intervention time $t$ is given by 
$$\widehat p_{i}(t)= \frac{\exp\{(H_i(t), X_i)^T\widehat\beta\}}{1+\exp\{(H_i(t), X_i)^T\widehat\beta\}}.$$

In the second step,  using the estimated propensity scores, according to (4) for $t\in {\cal T}$ and by the definition of the average intervention effect $\gamma(\Delta)$,  we estimate $\gamma(\Delta)$ explicitly as
$$\hspace{-2in}\widehat \gamma(\Delta)=\frac{\sum_{i=1}^{n}\int I(T_i=t)/\widehat p_i(t) Y_i(t+\Delta) d\widehat F_T(t)}{\sum_{i=1}^n \int I(T_i=t)/\widehat p_i(t) d\widehat F_T(t)}$$
$$
-\frac{\sum_{i=1}^{n}\int I(T_i>t+\Delta)/(1-\widehat p_i(t)) Y_i(t+\Delta) d\widehat F_T(t)}{\sum_{i=1}^n \int I(T_i>t+\Delta)/(1-\widehat p_i(t)) d\widehat F_T(t)},$$
where for the convenience of notation, we use $Y(t+\Delta)$ to denote $Y(t+\Delta; t)$ in subsequent exposition.
Removing the denominators in the above expression does not necessarily invalidate the consistency of the estimator, but can lead to substantial efficiency gain as shown in survey sampling literature (e.g., using standardized weights may improve efficiency). Specifically, let $t_{j(i)}$ be the intervention day for state $i$, let $X_{i,j(i)}=(H_i(t_{j(i)}), X_i)^T$, and define $\widehat p_i=\frac{\exp\{X_{i, j(i)}^T\widehat\beta\}}{1+\exp\{X_{i, j(i)}^T\widehat\beta\}}.$
Then in the second step, we estimate $\gamma(\Delta)$ by 
$$\widehat \gamma(\Delta)=\frac{\sum_{i=1}^{n}\sum_{j\in S(i)}d_{ij}\delta_{ij}/\widehat q_{ij}}{\sum_{i=1}^{n}\sum_{j\in S(i)}\delta_{ij}/\widehat q_{ij}}-\frac{\sum_{i=1}^{n}\sum_{j\in S(i)}d_{ij}(1-\delta_{ij})/(1-\widehat q_{ij})}{\sum_{i=1}^{n}\sum_{j\in S(i)}(1-\delta_{ij})/(1-\widehat q_{ij})},$$
where $d_{ij}$  is the change in reproduction number (i.e., $Y_i(j(i)+\Delta)$, or $R_{i,j(i)+\Delta}-R_{i,j(i)}$), $\delta_{ij}$  is the change in intervention status at time $j$ for state $i$, and 
$$\widehat q_{ij}=\frac{\exp\{X_{ij}^T\widehat\beta\}}{1+\exp\{X_{ij}^T\widehat\beta\}}, \ \ i=1,..,n, {j\in S(i)}.$$
Note $\widehat p_i=\widehat q_{i,j(i)}$.

The following theorem gives the asymptotic distribution for $\widehat\gamma(\Delta).$

\noindent{\bf Theorem 1}. Under assumptions (a)-(c) and assuming that $(H_i(t), X)$ is linearly independent with positive probability for some $t$ in ${\cal T}$ and that $H(t)$ has a bounded total variation in ${\cal T}$, $\sqrt n (\widehat \gamma(\Delta)-\gamma(\Delta))$ converges to a mean-zero normal distribution.

The asymptotic variance in Theorem 1 is given in the proof in the Appendix. A consistent estimator for the variance can be given by a plug-in estimator. Specifically,  the proof of Theorem 1 implies that $\sqrt n (\widehat\beta-\beta)$ is asymptotically normal, where $\beta$ is the true parameter value in the propensity score model,  and the asymptotic  variance can be consistently estimated by
$(\sum_{i=1}^n V_iV_i^T)/n$, where
\begin{equation*}
V_i=\left[n^{-1}\sum_{i=1}^{n}\sum_{j\in S(i)}
X_{ij}X_{ij}^T \widehat q_{ij}(1-\widehat q_{ij})\right]^{-1}\left\{\sum_{j\in S(i)}
X_{ij} (\delta_{ij}-\widehat q_{ij})\right\}.
\end{equation*}
Finally, through the linear expansion given in the proof of Theorem 1,  
If we let
$$A_i=\sum_{j\in S(i)}d_{ij}\delta_{ij}/\widehat q_{ij}, \ \
B_i=\sum_{j\in S(i)}\delta_{ij}/\widehat q_{ij},$$
and
$$C_i=\sum_{j\in S(i)}d_{ij}(1-\delta_{ij})/(1-\widehat q_{ij}), \ \
D_i=\sum_{j\in S(i)}(1-\delta_{ij})/(1-\widehat q_{ij})$$
and $\bar A, \bar B, \bar C$ and $\bar D$ be their respective average values,
then the asymptotic variance for $\widehat\gamma(\Delta)$ can be estimated as $\widehat\sigma^2=n^{-2}\sum_{i=1}^{n}(U_{i}-\bar U)^2,$
where 
\begin{eqnarray*}
U_{i}&=&\frac{A_i}{n^{-1}\sum_{k=1}^{n}B_k}-\frac{C_i}{n^{-1}\sum_{k=1}^{n}D_k}-\frac{n^{-1}\sum_{k=1}^{n}A_k}{(n^{-1}\sum_{k=1}^{n}B_k)^2}B_i
+\frac{n^{-1}\sum_{k=1}^{n}C_k}{(n^{-1}\sum_{k=1}^{n}D_k)^2}D_i\\
& &-\left[\frac{\sum_{k=1}^{n}\sum_{j\in S(k)}d_{kj}\delta_{kj}(1-\widehat q_{kj})X_{kj}^T/\widehat q_{kj}}{\sum_{k=1}^{n}B_k}\right.\\
& &\left.\qquad\qquad
-\frac{\sum_{k=1}^{n}A_k}{(\sum_{k=1}^{n}B_k)^2}\left(\sum_{k=1}^{n}\sum_{j\in S(k)}\delta_{kj}(1-\widehat q_{kj})X_{kj}^T/\widehat q_{kj}\right) \right]V_i\\
& &-\left[\frac{\sum_{k=1}^{n}\sum_{j\in S(k)}d_{kj}(1-\delta_{kj})\widehat q_{kj}X_{kj}^T/(1-\widehat q_{kj})}{\sum_{k=1}^{n}D_k}\right.\\
& &\left.\qquad\qquad
-\frac{\sum_{k=1}^{n}C_k}{(\sum_{k=1}^{n}D_k)^2}\left(\sum_{k=1}^{n}\sum_{j\in S(k)}(1-\delta_{kj})\widehat q_{kj}X_{kj}^T/(1-\widehat q_{kj})\right)\right] V_i\\
& &+\frac{1}{\sum_{k=1}^n B_k}\frac{d_{ii}}{q_{ii}}-\frac{\sum_{k=1}^{n}A_k}{(\sum_{k=1}^{n}B_k)^2}\frac{1}{q_{ii}}\\
& &-\frac{\sum_{k=1}^n (1-\delta_{ki})/(1-q_{ki})d_{ki}}{\sum_{k=1}^n D_k}
+\frac{\sum_{k=1}^n C_k}{(\sum_{k=1}^n D_k)^2}(\sum_{k=1}^n(1- \delta_{ki})/(1-q_{ki})).
\end{eqnarray*}
Therefore, the $95\%$-confidence interval for the average intervention effect is
$[\widehat\gamma(\Delta)-1.96\sqrt{\widehat\sigma^2}, \widehat{\gamma}(\Delta)+1.96\sqrt{\widehat\sigma^2}]$.

\noindent\textbf{Remark 1}. Since we may have a small number of states with an NPI when fitting the propensity score, the model can be either saturated or overfitted when the dimension of $X_i$ and $H_i(t)$ increases. We perform a screening step to obtain a parsimonious model for estimating the propensity scores.

\noindent\textbf{Remark 2}. The estimand $\gamma(\Delta)$ depends on the window size, $\Delta$,  between the intervention time $t$ and  effect time $(t+\Delta)$. We can vary different window sizes so as to obtain the estimated intervention effects over days since the intervention. This can be useful to study how long it might take for an intervention to become effective. 

\subsection{Estimation of HTE by regression}

A similar procedure can be applied to study the effect in a subgroup of states which share similar characteristics of $Z_i$ and moderation effects of $Z_i$ (here $Z_i$ is a subset of $X_i$). To estimate which factors in $Z_i$ may moderate the intervention effect, we use a regression model by assuming 
$$E[Y_i^{(1)}(t+\Delta;t)-Y_i^{(0)}(t+\Delta;t)|Z_i]=\theta^TZ_i.$$
Thus, testing the significance of $\theta$  identifies significant factors that moderate  intervention effect, a.k.a, HTE, which may lead to precision public health policy that targets states with certain characteristics.

Specifically, the estimator for $\theta$ can be obtained by solving 
$$\sum_{i=1}^n Z_i\left[
\int \left\{Y_i(t+\Delta)\left(\frac{ I(T_i=t)}{\widehat p_i(t)} -\frac{I(T_i>t+\Delta)}{1-\widehat p_i(t)}\right)-\theta^TZ_i\right\}I(T_i\ge t) d\widehat F_T(t)\right]=0,$$
or equivalently,
$$\sum_{i=1}^n Z_i\left[\sum_{j\in S(i)}\left\{d_{ij}\left(\frac{\delta_{ij}}{\widehat q_{ij}}-\frac{1-\delta_{ij}}{1-\widehat q_{ij}}\right)-\theta^TZ_i\right\}\right]=0.
$$
When $Z_i=1$, the derived estimator is asymptotically equivalent to $\widehat\gamma(\Delta)$ studied before. Let $\widehat\theta$ denote the estimator. 
Our next theorem states the asymptotic covariance of $\widehat\theta$.

\noindent\textbf{Theorem 2}. Under the assumptions in Theorem 1, if we further assume $E[Z_iZ_i^T]$ is non-singular, 
it holds
\begin{eqnarray*}
& &\sqrt n(\widehat\theta-\theta)\\
&=& E\left[ZZ^T\int I(T\ge t)dF_T(t)\right]^{-1}\\
& &\qquad \times\left[Z\left(
\int \left\{Y(t+\Delta)\left(\frac{I(T=t)}{p(t)} -\frac{I(T>t+\Delta)}{1-p(t)}\right)-\theta^TZ\right\}I(T\ge t)dF_T(t)\right)\right.\\
& &\left.
-\widetilde 
E[\widetilde Z \int  \widetilde Y(t+\Delta)\left(I(\widetilde T=t)\frac{1-p(t)}{p(t)}+I(\widetilde T>t+\Delta)\frac{p(t)}{1-p(t)}\right)\right.\\
& &\qquad \times \left.(\widetilde H(t), \widetilde X)^TI(T\ge t)d F_T(t)]S_{\beta}\right]+o_p(1),
\end{eqnarray*}
where $\widetilde E[\cdot]$ denotes the expectation with respect to $\widetilde Y$ and $\widetilde T$, and $S_{\beta}$ is the influence function for $\widehat\beta$ given in the proof of Theorem 1. Consequently, $\sqrt n (\widehat\theta-\theta)$ converges weakly to a mean-zero normal distribution.

The proof for Theorem 2 uses the same linear expansion argument as in the proof for Theorem 1 so is omitted. As a result of Theorem 2, the variance for $\widehat \theta$ can be consistently estimated by the following sandwich estimator, $\widehat \Psi=\Sigma_1^{-1}\Sigma_2\Sigma_1^{-1},$
where 
$$\Sigma_1=\sum_{i=1}^n \sum_{j\in S(i)}Z_iZ_i^T$$ and $\Sigma_2=\sum_{i=1}^n W_iW_i^T$ with
\begin{eqnarray*}
W_i&=&Z_i\left[\sum_{j\in S(i)}\left\{d_{ij}\left(\frac{\delta_{ij}}{\widehat q_{ij}}-\frac{1-\delta_{ij}}{1-\widehat q_{ij}}\right)
-\widehat\theta^TZ_i\right\}\right]\\
& &-\left[n^{-1}\sum_{k=1}^{n}Z_k\sum_{j\in S(k)}d_{kj}X_{kj}^T\left(\delta_{kj}\frac{1-\widehat q_{kj}}{\widehat q_{kj}}+(1-\delta_{kj})\frac{\widehat q_{kj}}{1-\widehat q_{kj}}\right)\right] V_i.
\end{eqnarray*}
Therefore, to test whether the $l$th component of $\theta$ is zero at a significance level of $\alpha$, we reject the null if $|\widehat\theta_l|/\sqrt{\widehat \Psi_{ll}}$ is larger than the $(1-\alpha/2)$-quantile of the standard normal distribution, where $\widehat\theta_l$ is the $l$th component of $\widehat\theta$ and $\widehat \Psi_{ll}$ is the $l$th diagonal element of $\widehat \Psi$.

\section{Analysis of US COVID-19 Data}
\label{s:application}
Since the first reported case in Washington on January 22, 2020, COVID-19 spread rapidly across US, especially in the northeast. During mid-March to early April, states issued lockdown orders (physical distance closures) after the national emergency was declared on March 13, 2020. Large declines in the number of daily new reported cases and deaths were seen in April and May after lockdown orders. However, a second surge of COVID-19 arrived in June after reopening, primarily in the southern and western states. From November 2020 to early 2021, US has experienced a third surge of COVID-19 while the mass vaccination started to take place. 

\begin{figure}
    \centering
    \subfloat[Calendar dates of interventions]
    {\includegraphics[width=0.75\textwidth]{"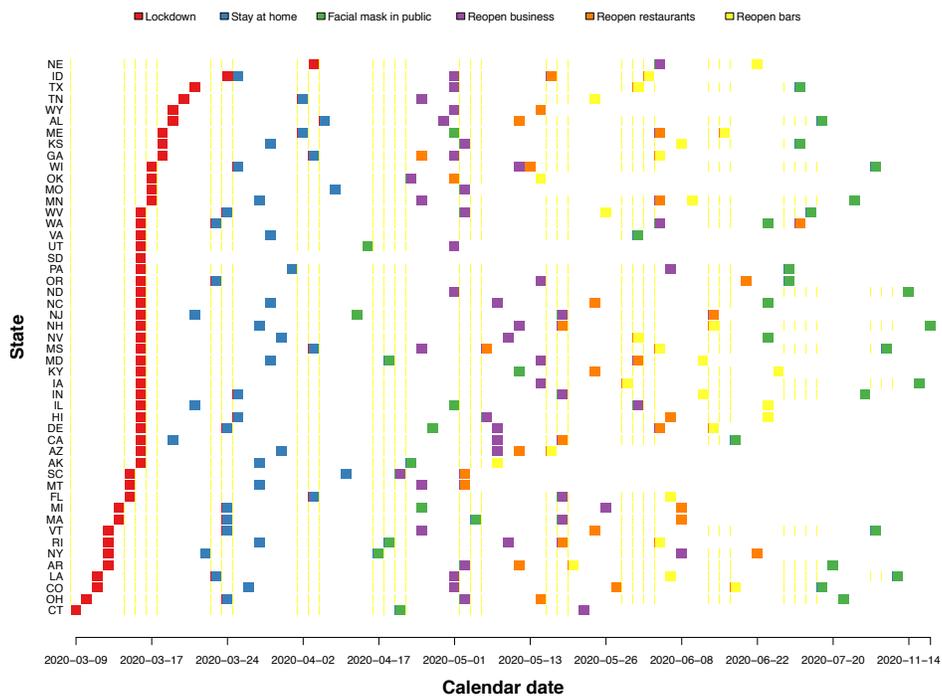"}\label{fig:policy_date}}\\\vskip .1in
    \subfloat[Intervention time]
    {\includegraphics[width=0.75\textwidth]{"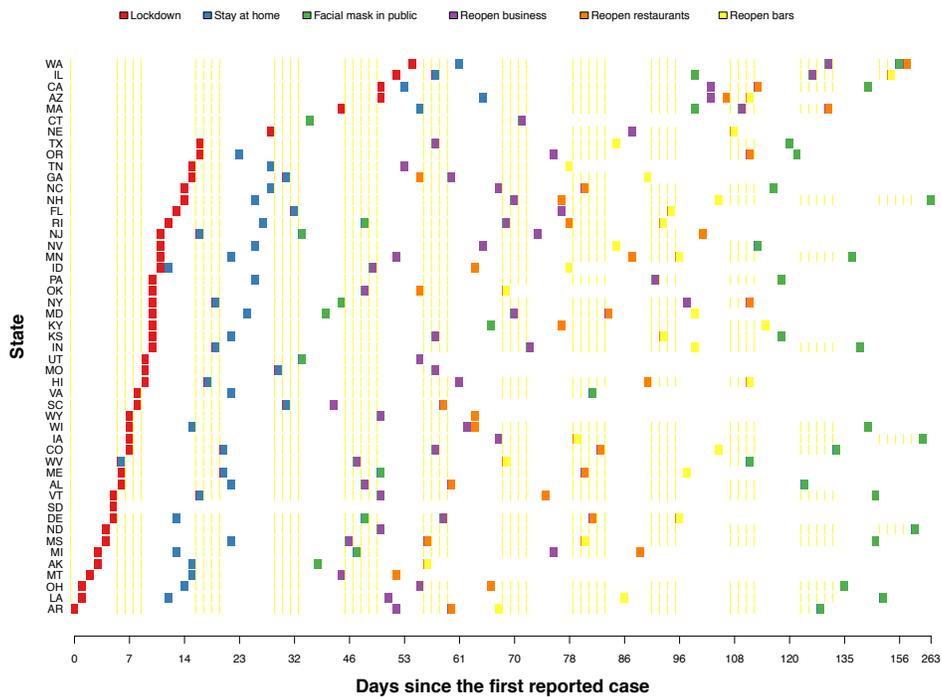"}\label{fig:policy_time}}\\
    \caption{Timing of interventions across states: calendar dates versus days since first reported case.}\label{fig:policy}
\end{figure}

We consider six state-wide NPIs: lockdown (date defined as the first physical distance closure), stay-at-home order, mandatory facial masks,  reopening business, reopening restaurants, and reopening bars. In our analysis, 48 states that have implemented an intervention after their first reported case were included. States issued lockdown orders between March 09 and April 3, 2020; 39 states placed stay-at-home order between March 19 and April 7; and 37 states mandated facial masks in public between April 8 and November 20. Between April 20 and June 8, 49 states issued reopening business order; 46 states issued reopen restaurant order between April 24 and July 3; and 44 states issued reopen bar order between May 1 and July 3. We  aligned  states by transforming calendar time to time since the first reported case. {Figure \ref{fig:policy} aligns states in two different ways: aligning by calendar dates (Figure \ref{fig:policy_date}), and aligning by  days since the first reported case (Figure \ref{fig:policy_time}). Two alignments differ, for example, many states implemented lockdown on March 16th but they were at different days since their first reported case. The latter alignment provides more variability between states and more meaningful measure as the stage in the pandemic.} Figure \ref{fig:policy_time} shows that stay-at-home order followed quickly after lockdown, and intervention times for other NPIs vary considerably across states. The intervention time of lockdown was between  (0, 54) days since the first reported case,  stay-at-home was between (6, 65) days, and  mandatory facial masks was between (34, 263) days. Reopening economy policies had a wider range of times between states. 
The gap time between implementing two different interventions also vary across states. We leverage these heterogeneity to match a ``case state"  with ``control states" without  interventions.


\begin{figure}
    \centering
    \includegraphics[width=\textwidth]{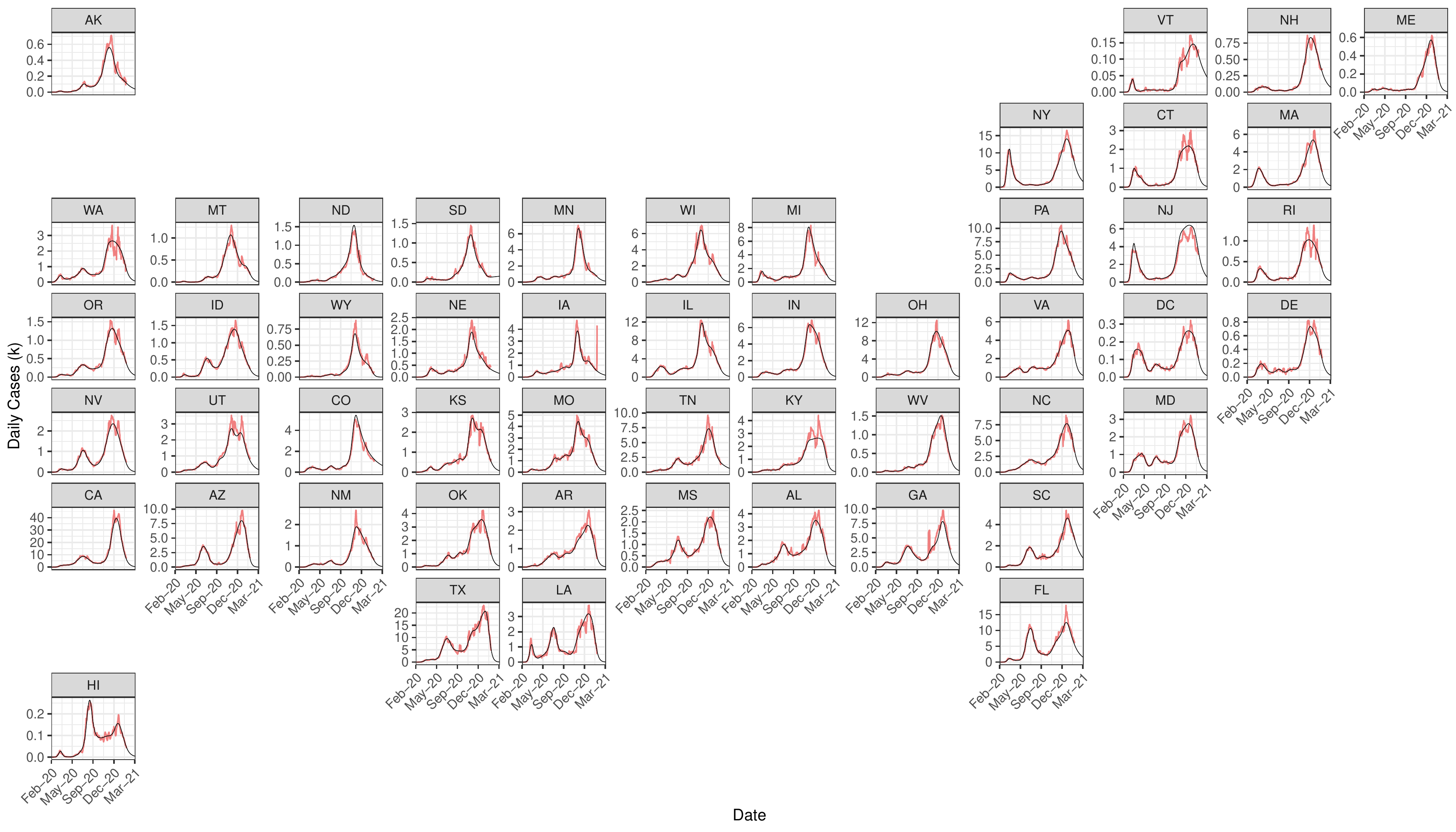}
    \caption{Observed (7-day moving average; red curve) and fitted (black curve) incidence COVID-19 cases from February 2020 to March 2021 in US States.}
    \label{fig:caseall}
\end{figure}

We fitted survival-convolution models for each state, using the daily incidence cases reported at {Johns Hopkins University Center for Systems Science and Engineering (JHU CSSE \citep{DONG2020533})} from the date of the first observed case as early as January 22, 2020 to February 16, 2021. This model successfully captured the epidemic trends of COVID-19 incidence cases in  50 states (Figure \ref{fig:caseall}). The fitted curves captured surges in large states such as New York, California, Florida, Texas, as well as smaller states including Maine, Wyoming, and the Dakotas. 
From the estimated new infections, we derive $R_t$ using equation (3). We show the estimated $R_t$ over the epidemic course in the Web Appendix Figure S1. 

\begin{figure}[!ht]
    \centering
    \includegraphics[width=1\textwidth]{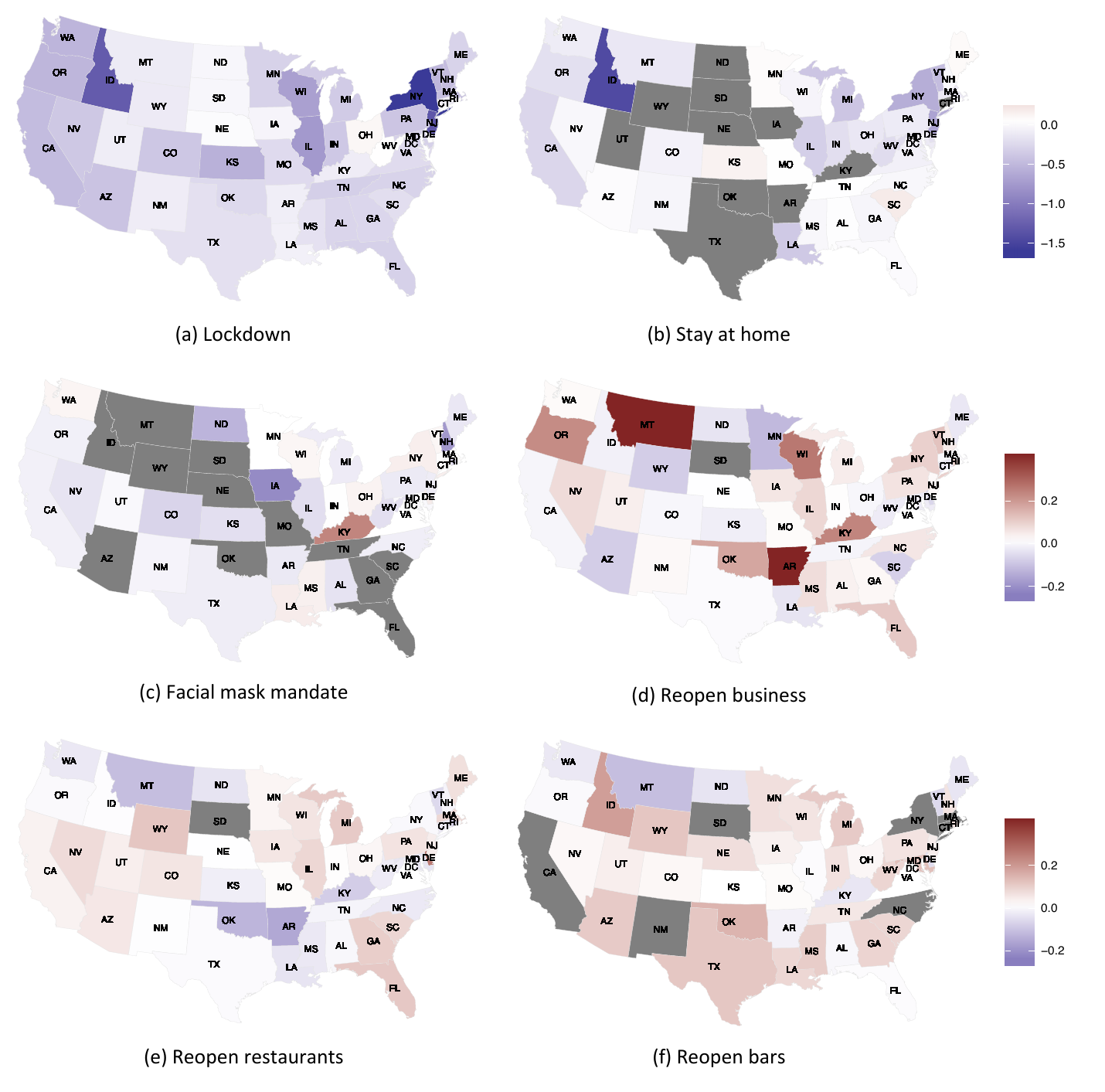}
    \vspace{20pt}
    \caption{Difference in $R_t$  between seven days post intervention and one day before intervention for each NPI in US States. Dark grey color indicates that a state had not implemented an NPI.}
    \label{fig:Rtcompare}
\end{figure}

To visualize observed changes in $R_t$ after each NPI, we present $R_t$ differences between seven days post intervention and one day before intervention in Figure \ref{fig:Rtcompare}. A darker cool color indicates a larger decrease in $R_t$ and a darker warm color indicates a larger increase. The states that did not implement certain NPIs are colored in gray. We see that $R_t$ in many states in the northeast and west decreased sharply 7 days after lockdown. For most states that had placed stay-at-home orders, $R_t$ also decreased after the orders. As a comparison, not all states showed a reduction in $R_t$ after facial mask mandates. Reopening business presents some degree of heterogeneity.  Among the three reopening interventions, reopening bars had the largest increase in $R_t$. These results show the observed changes in the states that had initiated NPIs, but lacks a control group. We will use the methods developed in Section 2 to formally estimated intervention effects  by  a DID estimator under the nested case-control design. 

Our goal is to formally quantify the impacts of NPIs and separate intervention effect from a natural decrease or increase trend in the absence of intervention using the inversely weighted DID estimator developed in Section 2.
We estimated the ATE $\gamma(\Delta)$, change in $R_t$ after $\Delta$ days of implementing the intervention. In our analysis, we evaluated lockdown's effect up to {6 days, stay-at-home orders up to 11 days}, and other interventions up to 14 days. Lockdown and stay-at-home orders had shorter evaluation period because they were enacted at relatively short time interval. A greater $\Delta$ would not satisfy assumption (a) or have enough eligible control states.  We regarded the states that had intervention time within $\Delta$ days as implementing the intervention at the same time. State-specific characteristics were included as covariates to construct propensity scores to account for differences between states. Given the associations between state-level characteristics and COVID-19 transmission and NPIs \citep{rader2020crowding,sy2020Socioeconomic,auger2020association}, the candidate covariates were the demographic characteristics including {the percentage of White, the percentage of Latino, the percentage of male, the percentage of age 65 and over, the percentage of male at age 65 and over}, CDC SVI variables \citep{cdcSVI} including {the percentage of below poverty, the percentage of unemployed, the percentage of no high school diploma, the percentage of speaking English ``less than well", the percentage of housing in structures with 10 or more units, the percentage of mobile homes, the percentage of more people than rooms at household level, the percentage of no vehicle, the percentage of in institutionalized group quarters, the percentage of civilian non-institutionalized population with a disability, the percentage of single parent households with children under 18, and per capita income}. The time-varying covariates including average $R_t$, average daily new reported cases, average daily new reported deaths, average rate of positive tests, and average {percentage of total inpatient beds utilized by patients who have probable or confirmed COVID-19 \citep{healthinpatient} during one week prior to the intervention. We standardized the unemployment variable by the state's population of aged 17-65, and standardized the other SVI variables except for per capita income by state’s total population. The time-varying covariates were also standardized by state's population and multiplied by 100,000.}  A different set of propensity scores was constructed for each $\Delta$ because eligible control states could change. We selected the top 10 covariates based on {Spearman rank correlation} for each intervention
separately and the covariates with a large proportion of  missing were excluded. 


Web Appendix Tables S1-S6 show the propensity score estimates of each intervention. The states with higher average pre-intervention $R_t$, larger average daily new cases, and larger average daily new deaths, {fewer persons who speak English "less than well"}, higher Latino population, higher institutionalized population, and {higher percentage of crowded household} were more likely to enact the lockdown order. For stay-at-home order, states with larger average daily new cases and smaller population of no high school diploma were more likely to implement this NPI. The states with larger average daily new cases were more likely to require wearing facial masks, and the states with {larger average daily new cases and deaths {and fewer mobile homes} were less likely to reopen bars.}

\begin{figure}[!ht]
\includegraphics[width=0.45\textwidth]{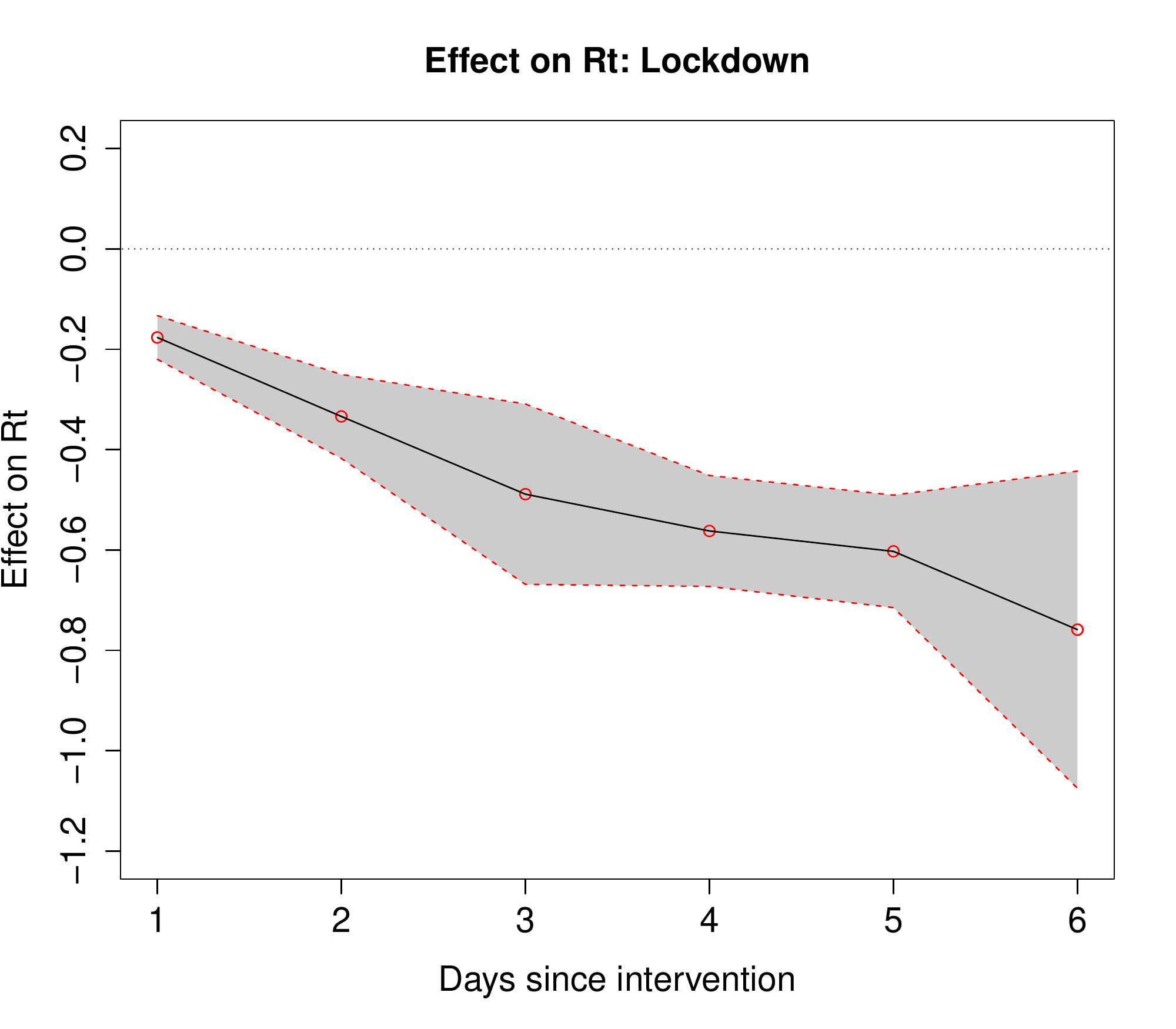}
\includegraphics[width=0.45\textwidth]{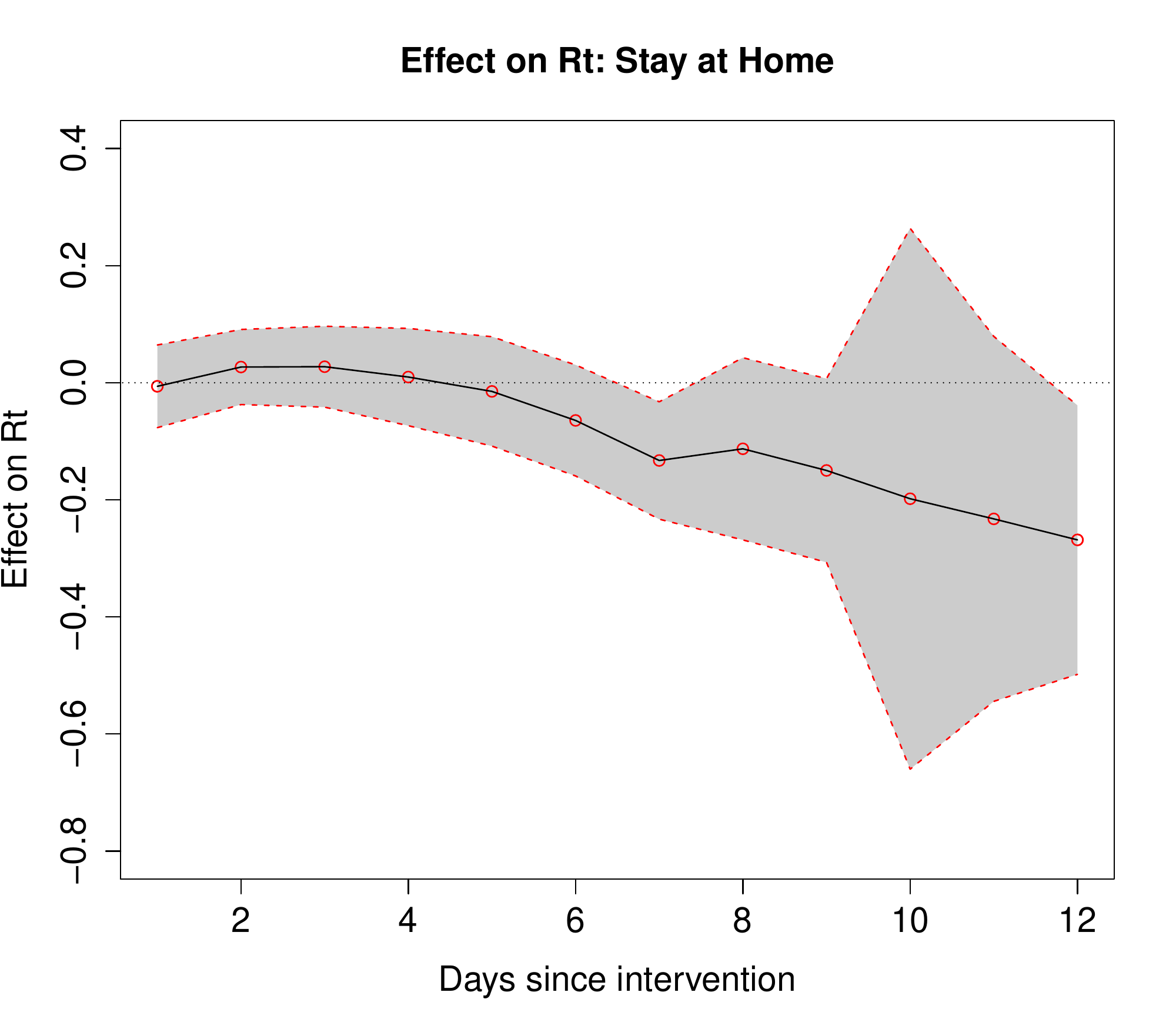} \vskip 0.1in
\includegraphics[width=0.45\textwidth]{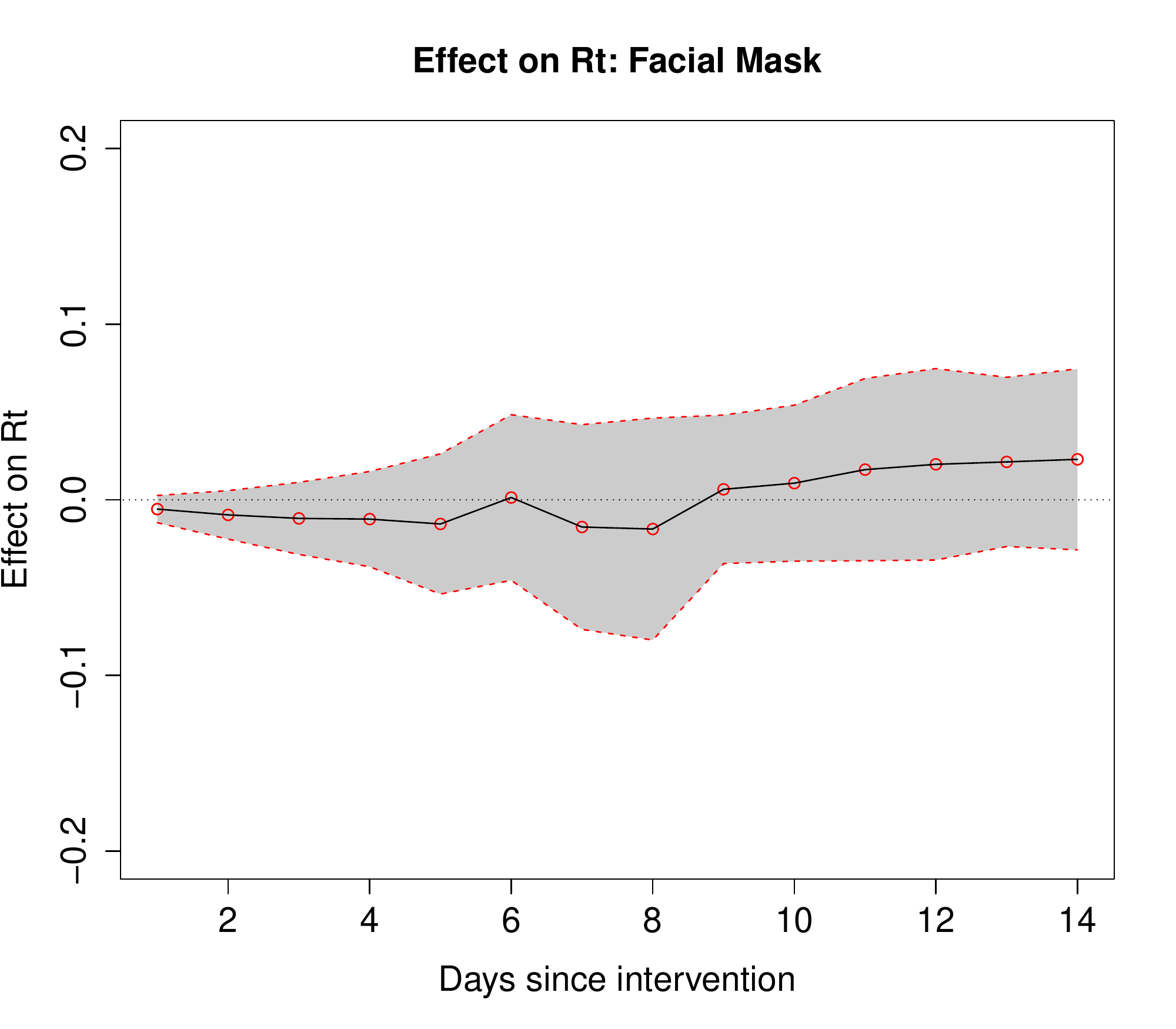} 
\includegraphics[width=0.45\textwidth]{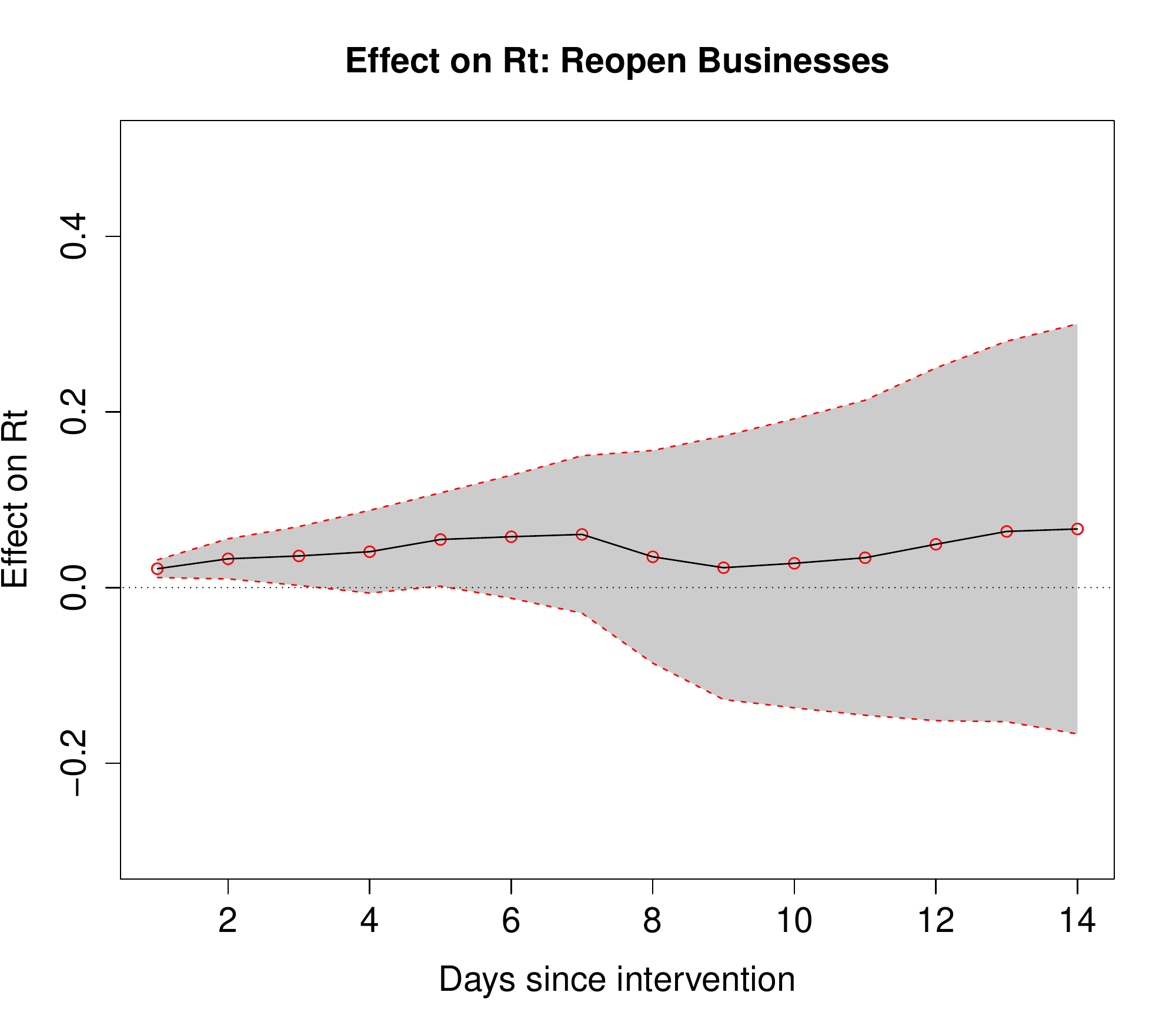}\vskip 0.1in
\includegraphics[width=0.45\textwidth]{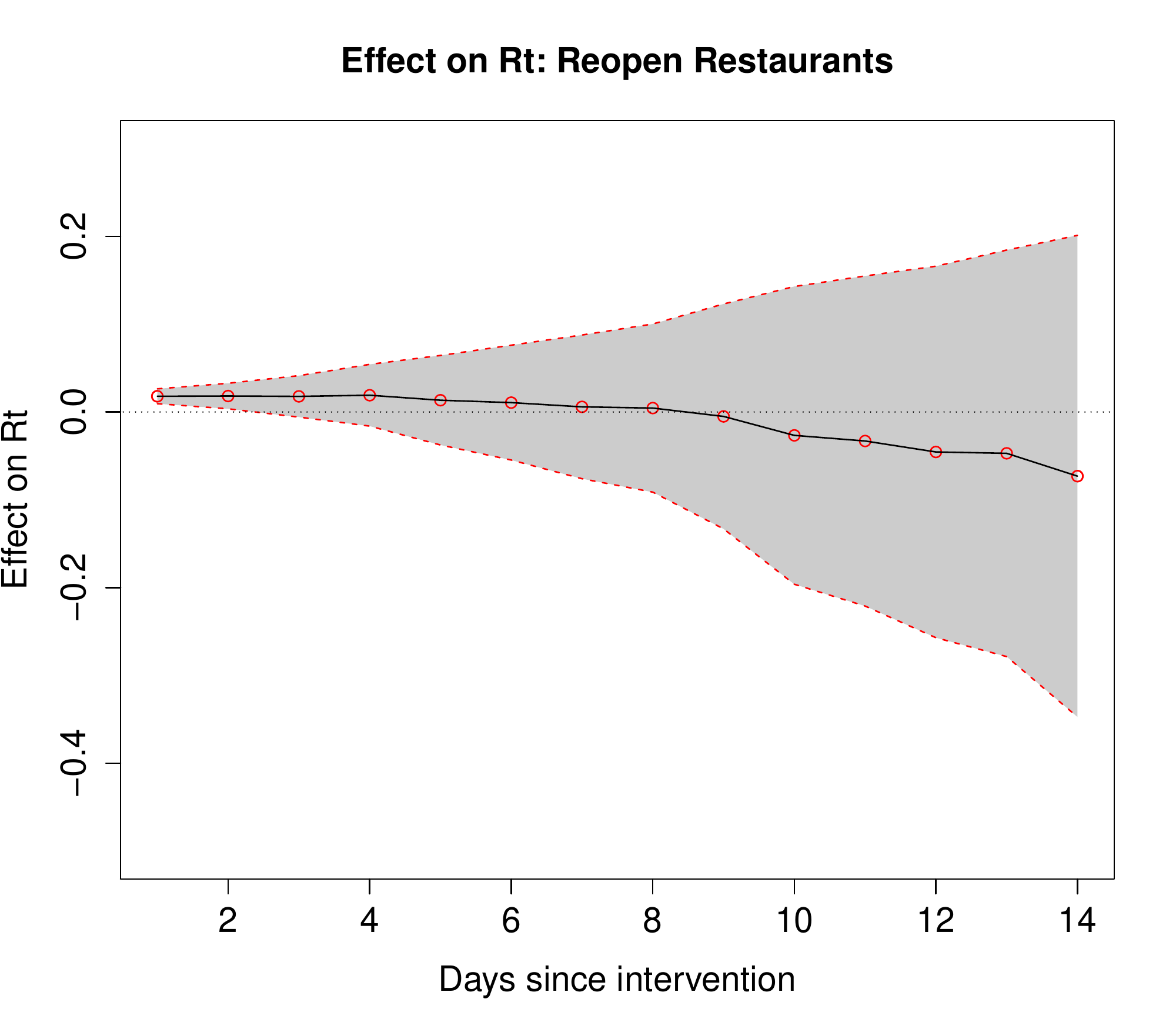}
\includegraphics[width=0.45\textwidth]{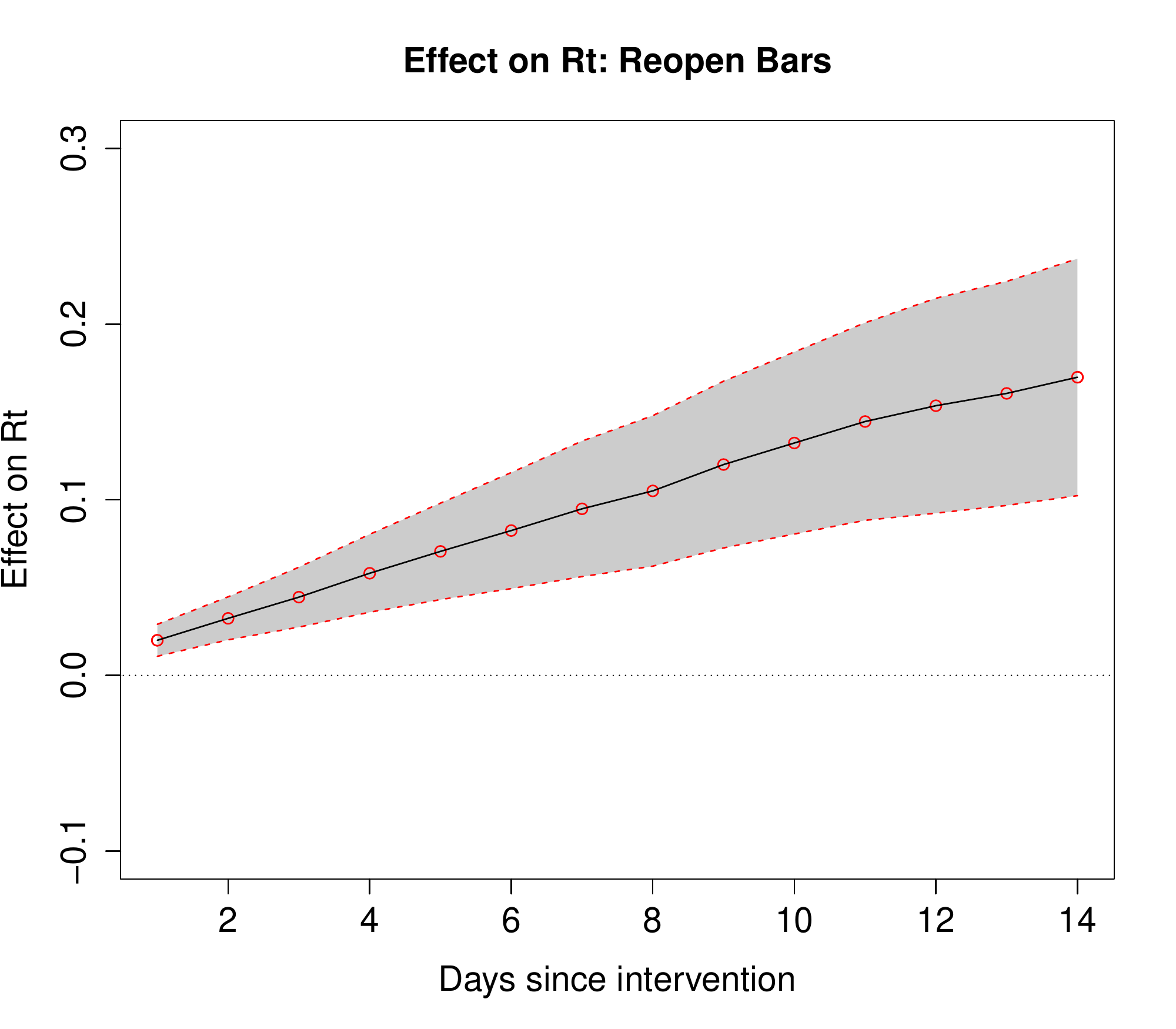}

\caption{Average intervention effects with 95\% confidence intervals.}\label{fig:ATE}
\end{figure}

 The ATEs of the six NPIs are shown in Table \ref{tab:ATE} and Figure \ref{fig:ATE}. Enacting lockdown significantly decreased $R_t$ immediately after its implementation,  {with an average effect of $-0.759$ (95\% CI, $-1.075$ to $-0.443$) six days after. The effect of stay-at-home order reached $-0.133$ (95\% CI, $-0.233$ to $-0.033$) seven days post-intervention.  Reopening bars  significantly increased $R_t$. The average effect of reopening bars was an increase of 0.095 (95\% CI, 0.056 to 0.134) after 7 days and reached 0.17 (95\% CI, 0.103 to 0.237) after 14 days. The ATE of reopening business was positive but not significant. The ATE of reopening restaurants and mask mandates was not significant.} 

\begin{table}[htp]
\small
\begin{center}
\caption{
\bf  {Average Intervention Effects of the Six NPIs}}\label{tab:ATE}
\resizebox{1\textwidth}{!}{
\begin{tabular}{lcccccc}
\toprule
&&& \multicolumn{1}{c}{{\bf Mask }} 
&\multicolumn{1}{c}{{\bf Reopen }} 
&\multicolumn{1}{c}{{\bf Reopen }} 
&\multicolumn{1}{c}{{\bf Reopen }}
\cr
\multicolumn{1}{l}{{\bf Day}} &\multicolumn{1}{c}{{\bf Lockdown}} 
&\multicolumn{1}{c}{{\bf Stay-at-home}} &\multicolumn{1}{c}{{\bf  Mandate}}
&\multicolumn{1}{c}{{\bf  Businesses}}
&\multicolumn{1}{c}{{\bf  Restaurants}} &\multicolumn{1}{c}{{\bf  Bars}}
  \cr
  	&	Estimate (se)		&	Estimate (se)		&	Estimate (se)		&	Estimate (se)		&	Estimate (se)		&	Estimate (se)		\cr
\midrule
		
$\Delta=1$	&	-0.176 (0.022)	&	-0.006 (0.036)	&	-0.005 (0.004)	&	0.022 (0.005)	&	0.018 (0.004)	&	0.020 (0.005)	\cr
$\Delta=2$	&	-0.334 (0.043)	&	0.027 (0.033)	&	-0.008 (0.007)	&	0.033 (0.012)	&	0.018 (0.007)	&	0.032 (0.006)	\cr
$\Delta=3$	&	-0.489 (0.092)	&	0.027 (0.035)	&	-0.010 (0.010)	&	0.036 (0.017)	&	0.018 (0.012)	&	0.044 (0.009)	\cr
$\Delta=4$	&	-0.562 (0.056)	&	0.010 (0.042)	&	-0.011 (0.014)	&	0.041 (0.024)	&	0.019 (0.018)	&	0.058 (0.011)	\cr
$\Delta=5$	&	-0.603 (0.057)	&	-0.015 (0.048)	&	-0.014 (0.020)	&	0.055 (0.027)	&	0.013 (0.026)	&	0.071 (0.014)	\cr
$\Delta=6$	&	-0.759 (0.161)	&	-0.064 (0.048)	&	0.001 (0.024)	&	0.058 (0.036)	&	0.011 (0.033)	&	0.082 (0.017)	\cr
$\Delta=7$	&	-	&	-0.133 (0.051)	&	-0.016 (0.030)	&	0.060 (0.046)	&	0.006 (0.042)	&	0.095 (0.020)	\cr
$\Delta=8$	&	-	&	-0.113 (0.079)	&	-0.017 (0.032)	&	0.035 (0.062)	&	0.004 (0.049)	&	0.105 (0.022)	\cr
$\Delta=9$	&	-	&	-0.150 (0.080)	&	0.006 (0.022)	&	0.023 (0.077)	&	-0.005 (0.065)	&	0.120 (0.024)	\cr
$\Delta=10$	&	-	&	-0.198 (0.236)	&	0.009 (0.023)	&	0.028 (0.084)	&	-0.027 (0.086)	&	0.132 (0.026)	\cr
$\Delta=11$	&	-	&	-0.233 (0.159)	&	0.017 (0.026)	&	0.034 (0.092)	&	-0.033 (0.096)	&	0.144 (0.029)	\cr
$\Delta=12$	&	-	&	-	&	0.020 (0.028)	&	0.049 (0.102)	&	-0.045 (0.108)	&	0.154 (0.031)	\cr
$\Delta=13$	&	-	&	-	&	0.022 (0.024)	&	0.064 (0.110)	&	-0.047 (0.118)	&	0.160 (0.032)	\cr
$\Delta=14$	&	-	&	-	&	0.023 (0.026)	&	0.067 (0.119)	&	-0.073 (0.140)	&	0.170 (0.034)	\cr
   \bottomrule
\end{tabular}
}
\end{center}
\normalsize{$-$ indicates the effect was not applicable at $\Delta$ day.}
\end{table}

We further assessed HTE to identify whether any factor moderates the intervention effects of lockdown, stay-at-home, and reopening policies. Our candidate moderators included the {percentage of age 65 and over, the percentage of White, the percentage of male, and the percentage below poverty}. {We did not find any significant moderator.} {The estimated HTE for race (percentage of White race) was marginally significant for reopening bars (Web Appendix Figure S2 shows the estimated HTE and  confidence interval of race  on reopening bars).}

\section{Discussion}
In this work, we propose a nested case-control design and propensity score weighting approach to evaluate impact of NPIs on mitigating COVID-19 transmission. Our method aligns states by transforming calendar time to time since the first reported case and allows each state to serve in both treated and control group during different time periods. Our  estimator provides causal intervention effect under assumptions and we further identify the factors that moderate intervention effect. Our analysis shows that mobility restricting policies (lockdown and stay-at-home orders) have a large effect on reducing transmission. The effect of mask mandate was not significant. However, this result should be interpreted with care because mask mandate may not directly increase the  adoption of mask wearing behavior in the public \citep{rader2021mask}. Using self reported mask wearing data may be more effective in evaluating the effect of masking.  Reopening {bars} had a significant effect on increasing transmission.

We investigated each intervention separately in this work and did not consider interaction between interventions given the sample size (50 states). To evaluate more detailed intervention packages and interaction between NPIs, county-level data can be useful to increase sample size. Assuming intervention effects to be additive, we can use the estimated treatment effect to determine the the optimal sequence of the treatment effects and timing for controlling disease outbreak. {Our assumptions might be violated if there are interference effects between neighboring states and there might be other potential confounders that are not adjusted for in the propensity score model.} When NPIs have delayed effect, methods developed for dynamic treatment regimes may be more appropriate.  As an extension, for county-level analysis we can borrow spatial information from counties that are similar and adjacent to each other to account for the transmission from region to region.  Other extensions to our method include using survival analysis to estimate the propensity scores for $T_i$ or adopting a doubly robust method to improve the IPW DID estimator.


\section*{Acknowledgments}
This research is supported by U.S. NIH grants GM124104, NS073671, and MH117458.

\newpage
\begin{appendices}
\section{Proof of Theorem 1}

We use ${\bf P}_n$ to denote the empirical measure associated with $n$ states' observations and use ${\bf P}$ to denote its expectation. 
First, using the estimating equation for $\beta$, we can easily show that 
$$\widehat\beta-\beta
=({\bf P}_n-{\bf P})S_{\beta}+o_p(n^{-1/2}),\eqno(A.1)
$$
where
$$S_{\beta}=\left[{\bf P}\int  (H(t), X)^T(H(t), X)I(T\ge t) p(t)(1-p(t))dF_T(t)\right]^{-1}$$
$$\times \int  (H(t), X)^TI(T\ge t) \left\{I(T=t)-p(t))\right\}dF_T(t),$$ 
and we note that the matrix inverse exists due to the linear independence assumption. Thus, $\sqrt n (\widehat\beta-\beta)$ converges to a mean-zero normal distribution with covariance matrix $E[S_{\beta}S_{\beta}^T]$.

Next, we rewrite $\widehat\gamma(\Delta)$ as
$$\widehat{\gamma}(\Delta)=
\frac{{\bf P}_n\int I(T=t)/\widehat p(t) Y(t+\Delta) d\widehat F_T(t)}{{\bf P}_n \int I(T=t)/\widehat p(t) d\widehat F_T(t)}
-\frac{{\bf P}_n\int I(T>t+\Delta)/(1-\widehat p(t)) Y(t+\Delta) d\widehat F_T(t)}{{\bf P}_n \int I(T>t+\Delta)/(1-\widehat p(t)) d\widehat F_T(t)}.$$
Using linear expansion and microscopic arguments, we obtain
\begin{eqnarray*}
& &\widehat\gamma(\Delta)-\gamma(\Delta) \\
&=&\frac{({\bf P}_n-{\bf P})\int I(T=t)/\widehat p(t) Y(t+\Delta) d\widehat F_T(t)}{{\bf P}_n \int I(T=t)/\widehat p(t) d\widehat F_T(t)}\\
& &
-\frac{{\bf P}\int I(T=t)/\widehat p(t) Y(t+\Delta) d\widehat F_T(t)}{{\bf P}_n \int I(T=t)/\widehat p(t) d\widehat F_T(t)
{\bf P} \int I(T=t)/\widehat p(t) d\widehat F_T(t)}({\bf P}_n-{\bf P})\int I(T=t)/\widehat p(t) d\widehat F_T(t)\\
& &-\frac{({\bf P}_n-{\bf P})\int I(T>t+\Delta)/(1-\widehat p(t)) Y(t+\Delta) d\widehat F_T(t)}{{\bf P}_n \int I(T>t+\Delta)/(1-\widehat p(t)) d\widehat F_T(t)}\\
& &+\frac{{\bf P}\int I(T>t+\Delta)/(1-\widehat p(t)) Y(t+\Delta) d\widehat F_T(t)}{{\bf P}_n \int I(T>t+\Delta)/(1-\widehat p(t)) d\widehat F_T(t){\bf P} \int I(T>t+\Delta)/(1-\widehat p(t)) d\widehat F_T(t)}\\
& & \qquad \qquad \times ({\bf P}_n-{\bf P}) \int I(T>t+\Delta)/(1-\widehat p(t)) d\widehat F_T(t)\\
& &+\frac{{\bf P}\int I(T=t)/\widehat p(t) Y(t+\Delta) d\widehat F_T(t)}{{\bf P} \int I(T=t)/\widehat p(t) d\widehat F_T(t)}
-\frac{{\bf P}\int I(T>t+\Delta)/(1-\widehat p(t)) Y(t+\Delta) d\widehat F_T(t)}{{\bf P} \int I(T>t+\Delta)/(1-\widehat p(t)) d\widehat F_T(t)}-\gamma(\Delta).
\end{eqnarray*}
On the other hand, based on assumptions (a)-(c), using the same argument in Section 2, we know
$$\gamma(\Delta)=
\frac{{\bf P}\int I(T=t)/ p(t) Y(t+\Delta) dF_T(t)}{{\bf P} \int I(T=t)/ p(t) dF_T(t)}
-\frac{{\bf P}\int I(T>t+\Delta)/(1-p(t)) Y(t+\Delta) d F_T(t)}{{\bf P} \int I(T>t+\Delta)/(1- p(t)) dF_T(t)}.$$
Thus, the last term in the expansion of $\widehat\gamma(\Delta)-\gamma(\Delta)$ can be further expanded as a linear functional of $(\widehat\beta-\beta)$ and $(\widehat F-F)$, where 
we further plug in the expansion in (A.1) and note that $(\widehat F-F_T)(t)=({\bf P}_n-{\bf P})I(T\le t)$.

Finally, since $I(T\ge t), \widehat p(t), \widehat F_T(t)$ and $Y(t)$ have bounded total variations so they are $P$-Donsker, we conclude 
$$\widehat\gamma(\Delta)-\gamma(\Delta)=({\bf P}_n-{\bf P}) \Gamma+o_p(n^{-1/2}),$$
where if we define
$$A=\int I(T=t)/ p(t) Y(t+\Delta) c F_T(t), \  \ B=\int I(T=t)/ p(t) d F_T(t)$$
and
$$C=\int I(T>t+\Delta)/(1-p(t)) Y(t+\Delta) d F_T(t),  \  \ D=\int I(T>t+\Delta)/(1-p(t))  d F_T(t),$$
\begin{eqnarray*}
\Gamma
&=&\frac{A}{E[B]}-\frac{C}{E[D]}-\frac{E[A]}{E[B]^2}\int \frac{I(T=t)}{p(t)} dF_T(t)+\frac{E[C]}{E[D]^2} \int \frac{I(T>t+\Delta)}{1- p(t)} d F_T(t)\\
& & +\frac{\widetilde E[\frac{I(\widetilde T=T)}{p(T)}\widetilde Y(T+\Delta)]}{E[B]}-
\frac{E[A]}{E[B]^2} \widetilde E[\frac{I(\widetilde T=T)}{p(T)}]\\
& &-\frac{\widetilde E[\frac{I(\widetilde T>T+\Delta)}{1-p(T)}\widetilde Y(T+\Delta)]}{E[D]}
+\frac{E[C]}{E[D]^2} \widetilde E[\frac{I(\widetilde T>T+\Delta)}{1-p(T)}]\\
& &-
\left[\frac{{\bf P}\int I(T=t)\frac{1-p(t)}{p(t)} (H(t), X)^TY(t+\Delta) d F_T(t)}{{\bf P} \int \frac{I(T=t)}{ p(t)} dF_T(t)}\right.\\
& &\qquad\left.
-\frac{{\bf P}\int \frac{I(T=t)}{p(t)} Y(t+\Delta) dF_T(t){\bf P} \int I(T=t)\frac{1-p(t)}{p(t)}  (H(t), X)^Td F_T(t)}{({\bf P} \int I(T=t)/ p(t) dF_T(t))^2}\right]S_{\beta}\\
& &
-\left[\frac{{\bf P}\int I(T>t+\Delta)\frac{p(t)}{1- p(t)}(H(t), X)^T Y(t+\Delta) d F_T(t)}{{\bf P} \int \frac{I(T>t+\Delta)}{1-p(t)} dF_T(t)}\right.\\
& &\qquad\left.
-\frac{{\bf P}\int \frac{I(T>t+\Delta)}{1- p(t)} Y(t+\Delta) d F_T(t){\bf P} \int I(T>t+\Delta)\frac{p(t)}{1-p(t)}(H(t), X)^T dF_T(t)}{({\bf P} \int \frac{I(T>t+\Delta)}{1-p(t)} dF_T(t))^2}\right]S_{\beta}.
\end{eqnarray*}
Here, $\widetilde E[\cdot]$ denotes the expectation with $\widetilde T$ and $\widetilde Y$.
Therefore, $\sqrt n (\widehat\gamma(\Delta)-\gamma(\Delta))$ converges to a mean-zero normal distribution with variance $E[\Gamma\Gamma^T]$.

\end{appendices}

\nocite{*}

\bibliographystyle{apalike}
\bibliography{ms_arxiv}

\newcommand{\noop}[1]{}
\begin{thebibliography}{}

\bibitem[Abadie et~al., 2010]{abadie2010synthetic}
Abadie, A., Diamond, A., and Hainmueller, J. (2010).
\newblock Synthetic control methods for comparative case studies: Estimating
  the effect of california’s tobacco control program.
\newblock {\em Journal of the American Statistical Association},
  105(490):493--505.

\bibitem[Auger et~al., 2020]{auger2020association}
Auger, K.~A., Shah, S.~S., Richardson, T., Hartley, D., Hall, M., Warniment,
  A., Timmons, K., Bosse, D., Ferris, S.~A., Brady, P.~W., Schondelmeyer,
  A.~C., and Thomson, J.~E. (2020).
\newblock Association between statewide school closure and {COVID}-19 incidence
  and mortality in the {US}.
\newblock {\em JAMA}, 324(9):859--870.

\bibitem[CDC, 2020]{cdcSVI}
CDC (2020).
\newblock Social vulnerability index.
\newblock {\em https://svi.cdc.gov}.

\bibitem[Cho, 2020]{cho2020quantifying}
Cho, S.-W. (2020).
\newblock Quantifying the impact of nonpharmaceutical interventions during the
  {COVID}-19 outbreak: The case of {S}weden.
\newblock {\em The Econometrics Journal}, 23(3):323--344.

\bibitem[Cori et~al., 2013]{cori2013new}
Cori, A., Ferguson, N.~M., Fraser, C., and Cauchemez, S. (2013).
\newblock A new framework and software to estimate time-varying reproduction
  numbers during epidemics.
\newblock {\em American Journal of Epidemiology}, 178(9):1505--1512.

\bibitem[Davies et~al., 2020]{davies2020effects}
Davies, N.~G., Kucharski, A.~J., Eggo, R.~M., Gimma, A., Edmunds, W.~J.,
  Jombart, T., O'Reilly, K., Endo, A., Hellewell, J., Nightingale, E.~S.,
  et~al. (2020).
\newblock Effects of non-pharmaceutical interventions on {COVID}-19 cases,
  deaths, and demand for hospital services in the uk: a modelling study.
\newblock {\em The Lancet Public Health}, 5(7):e375--e385.

\bibitem[Dong et~al., 2020]{DONG2020533}
Dong, E., Du, H., and Gardner, L. (2020).
\newblock An interactive web-based dashboard to track covid-19 in real time.
\newblock {\em The Lancet Infectious Diseases}, 20(5):533--534.

\bibitem[Ernster, 1994]{ernster1994nest}
Ernster, V. (1994).
\newblock Nested case-control studies.
\newblock {\em Preventive Medicine}, 23(5):587 -- 590.

\bibitem[Ferguson et~al., 2020]{ferguson2020impact}
Ferguson, N.~M., Laydon, D., Nedjati-Gilani, G., Imai, N., Ainslie, K.,
  Baguelin, M., Bhatia, S., Boonyasiri, A., Cucunub{\'a}, Z., Cuomo-Dannenburg,
  G., et~al. (2020).
\newblock Impact of non-pharmaceutical interventions ({NPI}s) to reduce
  {COVID}-19 mortality and healthcare demand.
\newblock {\em {Imperial College COVID-19 Response Team}}.

\bibitem[Flaxman et~al., 2020]{flaxman2020estimating}
Flaxman, S., Mishra, S., Gandy, A., Unwin, H. J.~T., Mellan, T.~A., Coupland,
  H., Whittaker, C., Zhu, H., Berah, T., Eaton, J.~W., et~al. (2020).
\newblock Estimating the effects of non-pharmaceutical interventions on
  {COVID}-19 in {E}urope.
\newblock {\em Nature}, 584(7820):257--261.

\bibitem[Hahn et~al., 2001]{hahn2001identification}
Hahn, J., Todd, P., and Van~der Klaauw, W. (2001).
\newblock Identification and estimation of treatment effects with a
  regression-discontinuity design.
\newblock {\em Econometrica}, 69(1):201--209.

\bibitem[HealthData, 2020]{healthinpatient}
HealthData (2020).
\newblock {COVID-19} reported patient impact and hospital capacity by state.
\newblock {\em https://healthdata.gov}.

\bibitem[Leatherdale, 2019]{leatherdale2019natural}
Leatherdale, S.~T. (2019).
\newblock Natural experiment methodology for research: a review of how
  different methods can support real-world research.
\newblock {\em International Journal of Social Research Methodology},
  22(1):19--35.

\bibitem[Li et~al., 2020]{li2020early}
Li, Q., Guan, X., Wu, P., Wang, X., Zhou, L., Tong, Y., Ren, R., Leung, K.~S.,
  Lau, E.~H., Wong, J.~Y., Xing, X., Xiang, N., Wu, Y., Li, C., Chen, Q., Li,
  D., Liu, T., Zhao, J., Liu, M., Tu, W., Chen, C., Jin, L., Yang, R., Wang,
  Q., Zhou, S., Wang, R., Liu, H., Luo, Y., Liu, Y., Shao, G., Li, H., Tao, Z.,
  Yang, Y., Deng, Z., Liu, B., Ma, Z., Zhang, Y., Shi, G., Lam, T.~T., Wu,
  J.~T., Gao, G.~F., Cowling, B.~J., Yang, B., Leung, G.~M., and Feng, Z.
  (2020).
\newblock Early transmission dynamics in {Wuhan, China}, of novel
  {Coronavirus–Infected Pneumonia}.
\newblock {\em New England Journal of Medicine}, 382(13):1199--1207.
\newblock PMID: 31995857.

\bibitem[Nishiura et~al., 2020]{nishiura2020serial}
Nishiura, H., Linton, N.~M., and Akhmetzhanov, A.~R. (2020).
\newblock Serial interval of novel coronavirus ({COVID}-19) infections.
\newblock {\em International Journal of Infectious Diseases}, 93:284--286.

\bibitem[Oran and Topol, 2020]{oran2020prevalence}
Oran, D.~P. and Topol, E.~J. (2020).
\newblock Prevalence of asymptomatic {SARS-CoV-2} infection: a narrative
  review.
\newblock {\em Annals of Internal Medicine}, 173(5):362--367.

\bibitem[Pei et~al., 2020]{Pei2020Differential}
Pei, S., Kandula, S., and Shaman, J. (2020).
\newblock Differential effects of intervention timing on {COVID-19} spread in
  the {United States}.
\newblock {\em Science Advances}, 6(49).

\bibitem[Rader et~al., 2020a]{rader2020geographic}
Rader, B., Astley, C.~M., Sy, K. T.~L., Sewalk, K., Hswen, Y., Brownstein,
  J.~S., and Kraemer, M. U.~G. (2020a).
\newblock Geographic access to {United States SARS-CoV-2} testing sites
  highlights healthcare disparities and may bias transmission estimates.
\newblock {\em Journal of Travel Medicine}, 27(7).

\bibitem[Rader et~al., 2020b]{rader2020crowding}
Rader, B., Scarpino, S.~V., Nande, A., Hill, A.~L., Adlam, B., Reiner, R.~C.,
  Pigott, D.~M., Gutierrez, B., Zarebski, A.~E., Shrestha, M., et~al. (2020b).
\newblock Crowding and the shape of {COVID-19} epidemics.
\newblock {\em Nature Medicine}, 26(12):1829--1834.

\bibitem[Rader et~al., 2021]{rader2021mask}
Rader, B., White, L.~F., Burns, M.~R., Chen, J., Brilliant, J., Cohen, J.,
  Shaman, J., Brilliant, L., Kraemer, M. U.~G., Hawkins, J.~B., Scarpino,
  S.~V., Astley, C.~M., and Brownstein, J.~S. (2021).
\newblock Mask-wearing and control of {SARS-CoV}-2 transmission in the {USA}: a
  cross-sectional study.
\newblock {\em The Lancet Digital Health}, 3(3):e148--e157.

\bibitem[Ray et~al., 2020]{Ray2020medrxiv}
Ray, E.~L., Wattanachit, N., Niemi, J., Kanji, A.~H., House, K., Cramer, E.~Y.,
  Bracher, J., Zheng, A., Yamana, T.~K., Xiong, X., Woody, S., Wang, Y., Wang,
  L., Walraven, R.~L., Tomar, V., Sherratt, K., Sheldon, D., Reiner, R.~C.,
  Prakash, B.~A., Osthus, D., Li, M.~L., Lee, E.~C., Koyluoglu, U., Keskinocak,
  P., Gu, Y., Gu, Q., George, G.~E., Espa{\~n}a, G., Corsetti, S., Chhatwal,
  J., Cavany, S., Biegel, H., Ben-Nun, M., Walker, J., Slayton, R., Lopez, V.,
  Biggerstaff, M., Johansson, M.~A., Reich, N.~G., and et~al. (2020).
\newblock Ensemble forecasts of {Coronavirus disease} 2019 ({COVID-19}) in the
  {U.S.}
\newblock {\em medRxiv}.

\bibitem[Scire et~al., 2020]{scire2020reproductive}
Scire, J., Nadeau, S.~A., Vaughan, T.~G., Gavin, B., Fuchs, S., Sommer, J.,
  Koch, K.~N., Misteli, R., Mundorff, L., G{\"o}tz, T., et~al. (2020).
\newblock Reproductive number of the {COVID}-19 epidemic in {S}witzerland with
  a focus on the cantons of basel-stadt and basel-landschaft.
\newblock {\em Swiss Medical Weekly}, 150(19-20):w20271.

\bibitem[Sy et~al., 2020]{sy2020Socioeconomic}
Sy, K. T.~L., Martinez, M.~E., Rader, B., and White, L.~F. (2020).
\newblock {Socioeconomic disparities in subway use and COVID-19 outcomes in New
  York City}.
\newblock {\em American Journal of Epidemiology}.

\bibitem[Wang et~al., 2020]{wang2020survival}
Wang, Q., Xie, S., Wang, Y., and Zeng, D. (2020).
\newblock Survival-convolution models for predicting {COVID}-19 cases and
  assessing effects of mitigation strategies.
\newblock {\em Frontiers in Public Health}, 8:325.

\bibitem[Wing et~al., 2018]{wing2018designing}
Wing, C., Simon, K., and Bello-Gomez, R.~A. (2018).
\newblock Designing difference in difference studies: best practices for public
  health policy research.
\newblock {\em Annual Review of Public Health}, 39:453--469.

\end{thebibliography}
\clearpage

\newpage
\section*{Supporting information}
\renewcommand\thefigure{\thesection\arabic{figure}}
\setcounter{figure}{0}
\setcounter{figure}{0}
\renewcommand{\thefigure}{S\arabic{figure}}

\renewcommand\thetable{\thesection\arabic{table}}
\setcounter{table}{0}
\setcounter{table}{0}
\renewcommand{\thetable}{S\arabic{table}}

\subsection*{Web Appendix A: $R_t$ and Propensity Score Estimation}
\begin{figure}[!ht]
    \centering
    \includegraphics[width=\textwidth]{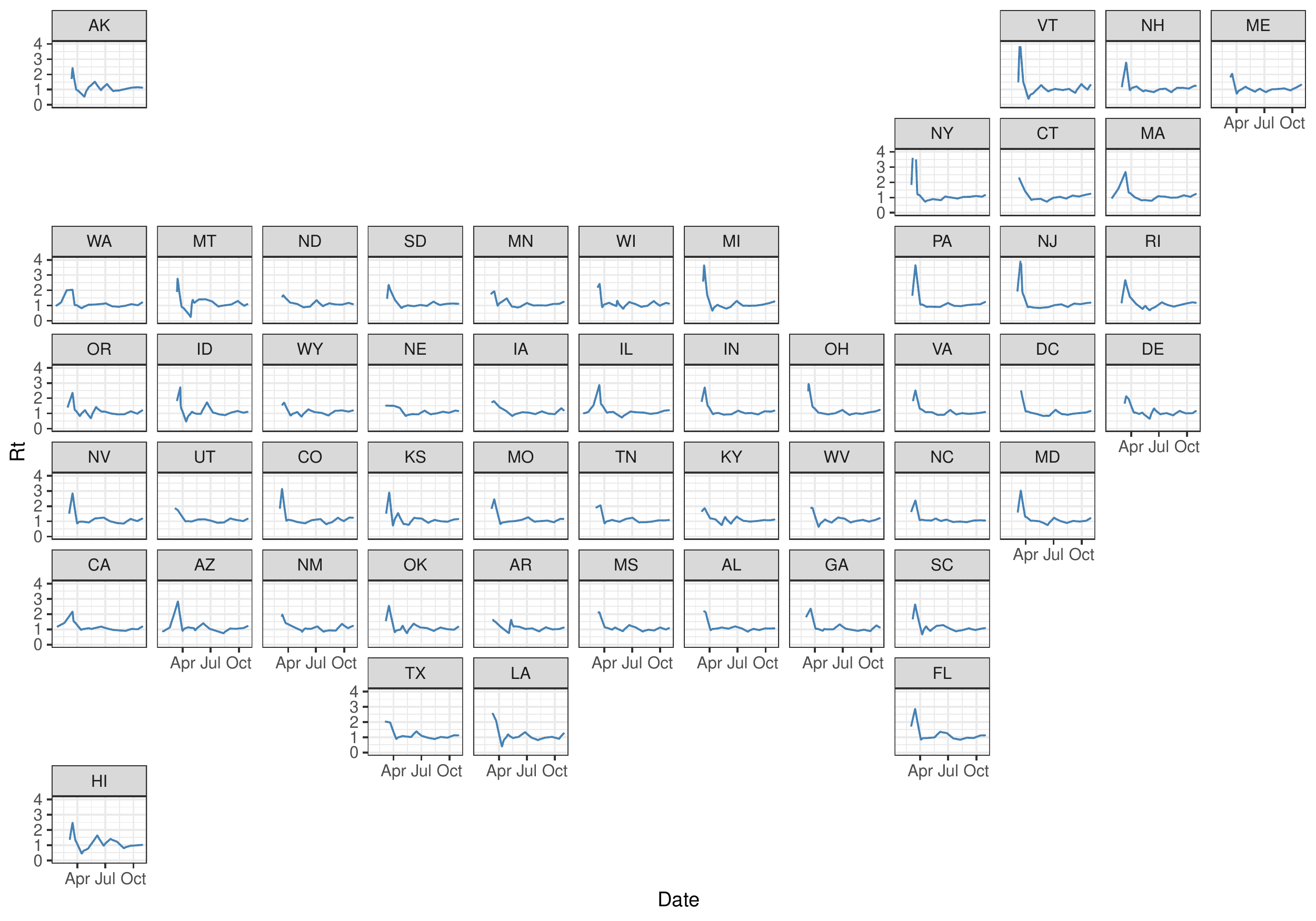}
    \caption{Estimated Effective Reproduction Number $R_t$ From February 2020 to February 2021 in the US}
    \label{fig:RtNPI}
\end{figure}

This section shows the results of estimated effective reproduction number $R_t$ in the US (Figure ~\ref{fig:RtNPI}), and the propensity score models for each intervention (Tables ~\ref{tab:prop_lockdown},  ~\ref{tab:prop_stay},  ~\ref{tab:prop_mask},   ~\ref{tab:prop_business},  ~\ref{tab:prop_restaurant}, and ~\ref{tab:prop_bar}). 

\begin{table}
\small
\centering
\caption{
\bf  {Propensity Score Estimates of Lockdown}}\label{tab:prop_lockdown}
\resizebox{1.05\textwidth}{!}{
\begin{tabular}{lcccccccc}
\toprule

&\multicolumn{1}{c}{{\bf New}} & &\multicolumn{1}{c}{{\bf New}}  &\multicolumn{1}{c}{{\bf Limit}} & &\multicolumn{1}{c}{\bf Multi-Unit } &&\multicolumn{1}{c}{\bf Crowded} 

\cr

\multicolumn{1}{l}{{\bf Day}}  &\multicolumn{1}{c}{{\bf Case}} &\multicolumn{1}{c}{\boldmath {$R_t$}} &\multicolumn{1}{c}{\bf Death} &\multicolumn{1}{c}{\bf English}  &\multicolumn{1}{c}{{\bf Latino}} &\multicolumn{1}{c}{{\bf House}} &\multicolumn{1}{c}{\bf  Institutionalized}  &\multicolumn{1}{c}{\bf Household} \cr

  	&	$\widehat\beta$ (p-value)		&	$\widehat\beta$ (p-value)		&	$\widehat\beta$ (p-value)		&	$\widehat\beta$ (p-value)		&	$\widehat\beta$ (p-value)		&	$\widehat\beta$ (p-value)		&	$\widehat\beta$ (p-value) 	&	$\widehat\beta$ (p-value) 		\cr
  	
\midrule
$\Delta=1$	&	8.144 (0)	&	3.982 (0)	&	-64.394 (0.023)	&	-1.172 (0.001)	&	0.183 (0.001)	&	-0.068 (0.706)	&	1.189 (0.002)	&	0.953 (0.242)	\cr
$\Delta=2$	&	11.901 (0)	&	5.085 (0)	&	-83.003 (0)	&	-1.493 (0.001)	&	0.220 (0)	&	-0.114 (0.599)	&	1.612 (0.003)	&	1.824 (0.120)	\cr
$\Delta=3$	&	15.526 (0)	&	6.863 (0)	&	-116.992 (0)	&	-1.739 (0.007)	&	0.247 (0.003)	&	-0.095 (0.794)	&	2.242 (0.001)	&	2.320 (0.063)	\cr
$\Delta=4$	&	28.865 (0)	&	8.230 (0)	&	31.493 (0.676)	&	-2.276 (0.002)	&	0.297 (0.003)	&	0.140 (0.632)	&	2.615 (0.018)	&	1.968 (0.203)	\cr
$\Delta=5$	&	39.656 (0)	&	12.747 (0)	&	-56.959 (0.407)	&	-2.846 (0.005)	&	0.365 (0.015)	&	0.291 (0.279)	&	3.300 (0.003)	&	2.412 (0.015)	\cr
$\Delta=6$	&	219.911 (0.107)	&	68.378 (0.09)	&	-558.381 (0.003)	&	-31.124 (0.104)	&	4.393 (0.102)	&	5.518 (0.062)	&	21.468 (0.153)	&	24.418 (0.226)	\cr
$\Delta=7$	&	-	&	-	&	-	&	-	&	-	&	-	&	-	&	-	\cr
$\Delta=8$	&	-	&	-	&	-	&	-	&	-	&	-	&	-	&	-	\cr
$\Delta=9$	&	-	&	-	&	-	&	-	&	-	&	-	&	-	&	-	\cr
$\Delta=10$	&	-	&	-	&	-	&	-	&	-	&	-	&	-	&	-	\cr
$\Delta=11$	&	-	&	-	&	-	&	-	&	-	&	-	&	-	&	-	\cr
$\Delta=12$	&	-	&	-	&	-	&	-	&	-	&	-	&	-	&	-	\cr
$\Delta=13$	&	-	&	-	&	-	&	-	&	-	&	-	&	-	&	-	\cr
$\Delta=14$	&	-	&	-	&	-	&	-	&	-	&	-	&	-	&	-	\cr
   \bottomrule
\end{tabular}
}
{\begin{flushleft} $-$ indicates the variable was not applicable at $\Delta$ day.
\end{flushleft}}
\end{table}

\begin{table}
\small
\centering
\caption{
\bf  {Propensity Score Estimates of Stay-at-home}}\label{tab:prop_stay}
\resizebox{1.1\textwidth}{!}{
\begin{tabular}{lccccccccccc}
\toprule

 &\multicolumn{1}{c}{{\bf New }} 
 &\multicolumn{1}{c}{\bf New} &\multicolumn{1}{c}{{\bf Limit}} & & &\multicolumn{1}{c}{\bf Multi-Unit}  &\multicolumn{1}{c}{\bf No High School} &\multicolumn{1}{c}{\bf Crowded} & & &
  \cr
  
  \multicolumn{1}{l}{{\bf Day}} &\multicolumn{1}{c}{{\bf Death}} 
 &\multicolumn{1}{c}{\bf Case} &\multicolumn{1}{c}{{\bf English}} &\multicolumn{1}{c}{\bf {Latino}} &\multicolumn{1}{c}{\boldmath {$R_t$}}  &\multicolumn{1}{c}{\bf House}  &\multicolumn{1}{c}{\bf Diploma} &\multicolumn{1}{c}{\bf Household} &\multicolumn{1}{c}{\bf Unemployed} &\multicolumn{1}{c}{\bf Institutionalized} &\multicolumn{1}{c}{\bf Disabled}
  \cr

  	&	$\widehat\beta$ (p-value)		&	$\widehat\beta$ (p-value)		&	$\widehat\beta$ (p-value)		&	$\widehat\beta$ (p-value)		&	$\widehat\beta$ (p-value)		&	$\widehat\beta$ (p-value)		&	$\widehat\beta$ (p-value) 	&	$\widehat\beta$ (p-value) 	&	$\widehat\beta$ (p-value) 	&	$\widehat\beta$ (p-value) &	$\widehat\beta$ (p-value)	\cr
  	
\midrule
$\Delta=1$	&	25.416 (0.007)	&	0.838 (0)	&	0.346 (0.279)	&	0.003 (0.941)	&	-0.201 (0.548)	&	-0.319 (0.169)	&	-0.618 (0.003)	&	0.473 (0.316)	&	-	&	-	&	-	\cr
$\Delta=2$	&	27.021 (0.013)	&	0.993 (0)	&	0.436 (0.221)	&	-0.002 (0.965)	&	-0.259 (0.464)	&	-0.401 (0.089)	&	-0.739 (0.001)	&	0.616 (0.183)	&	-	&	-	&	-	\cr
$\Delta=3$	&	31.587 (0.013)	&	1.168 (0)	&	0.609 (0.195)	&	0.002 (0.976)	&	-0.477 (0.241)	&	-0.519 (0.102)	&	-0.899 (0.003)	&	-	&	0.052 (0.934)	&	-	&	-	\cr
$\Delta=4$	&	40.988 (0.019)	&	1.398 (0)	&	0.621 (0.059)	&	-0.003 (0.941)	&	0.055 (0.884)	&	-0.563 (0.018)	&	-0.989 (0.002)	&	0.904 (0.079)	&	-	&	-	&	-	\cr
$\Delta=5$	&	48.896 (0.030)	&	1.746 (0.001)	&	0.836 (0.007)	&	-0.028 (0.513)	&	-	&	-0.612 (0.011)	&	-1.046 (0.001)	&	0.794 (0.159)	&	-	&	-0.155 (0.752)	&	-	\cr
$\Delta=6$	&	46.312 (0.102)	&	2.070 (0)	&	0.891 (0.002)	&	-0.039 (0.408)	&	-	&	-0.589 (0.013)	&	-1.112 (0.004)	&	0.883 (0.118)	&	-	&	-0.347 (0.466)	&	-	\cr
$\Delta=7$	&	29.690 (0.368)	&	3.021 (0)	&	0.743 (0.002)	&	-0.019 (0.728)	&	-	&	-0.575 (0.011)	&	-1.073 (0.006)	&	0.969 (0.073)	&	-	&	-0.340 (0.443)	&	-	\cr
$\Delta=8$	&	17.192 (0.597)	&	3.901 (0.001)	&	0.049 (0.911)	&	0.007 (0.918)	&	-	&	-0.281 (0.270)	&	-	&	1.996 (0.023)	&	-2.166 (0.001)	&	-0.194 (0.759)	&	-	\cr
$\Delta=9$	&	39.736 (0.314)	&	4.366 (0)	&	0.264 (0.640)	&	-0.079 (0.401)	&	-	&	-0.490 (0.161)	&	-	&	2.247 (0.003)	&	-1.943 (0.002)	&	-1.098 (0.063)	&	-	\cr
$\Delta=10$	&	42.246 (0.381)	&	6.320 (0)	&	0.184 (0.792)	&	0.009 (0.905)	&	-	&	-0.373 (0.228)	&	-	&	2.082 (0.009)	&	-2.168 (0.034)	&	-	&	0.350 (0.699)	\cr
$\Delta=11$	&	52.587 (0.410)	&	7.453 (0)	&	0.204 (0.706)	&	0.053 (0.418)	&	-	&	-0.225 (0.487)	&	-	&	2.187 (0.048)	&	-2.736 (0.017)	&	-	&	0.700 (0.445)	\cr
$\Delta=12$	&	63.706 (0.418)	&	10.495 (0.006)	&	-0.982 (0.351)	&	0.179 (0.217)	&	-	&	0.524 (0.400)	&	-	&	2.616 (0.321)	&	-1.929 (0.119)	&	-	&	1.158 (0.145)	\cr
$\Delta=13$	&	-	&	-	&	-	&	-	&	-	&	-	&	-	&	-	&	-	&	-	&	-	\cr
$\Delta=14$	&	-	&	-	&	-	&	-	&	-	&	-	&	-	&	-	&	-	&	-	&	-	\cr

   \bottomrule
\end{tabular}
}
{\begin{flushleft} $-$ indicates the variable was not selected or not applicable at $\Delta$ day.
\end{flushleft}}
\end{table}

\begin{table}
\small
\centering
\caption{
\bf  {Propensity Score Estimates of Mandatory Facial Mask}}\label{tab:prop_mask}
\resizebox{1.1\textwidth}{!}{
\begin{tabular}{lccccccccccc}
\toprule

 &\multicolumn{1}{c}{{\bf New}} 
& &\multicolumn{1}{c}{\bf New} &\multicolumn{1}{c}{\bf {No}} &\multicolumn{1}{c}{\bf {Limit}} &  & &\multicolumn{1}{c}{\bf {Mobile}} &\multicolumn{1}{c}{\bf { Male at Age 65}}
  \cr

  \multicolumn{1}{l}{{\bf Day}} &\multicolumn{1}{c}{{\bf Case}} 
&\multicolumn{1}{c}{\boldmath {$R_t$}} &\multicolumn{1}{c}{\bf Death} &\multicolumn{1}{c}{{\bf Vehicle}} &\multicolumn{1}{c}{\bf {English}} &\multicolumn{1}{c}{\bf Latino}  &\multicolumn{1}{c}{\bf Unemployed} &\multicolumn{1}{c}{\bf Home} &\multicolumn{1}{c}{\bf {and over}} &\multicolumn{1}{c}{\bf {Male}} &\multicolumn{1}{c}{\bf {White}}
  \cr

  	&	$\widehat\beta$ (p-value)		&	$\widehat\beta$ (p-value)		&	$\widehat\beta$ (p-value)		&	$\widehat\beta$ (p-value)		&	$\widehat\beta$ (p-value)		&	$\widehat\beta$ (p-value)		&	$\widehat\beta$ (p-value) 	&	$\widehat\beta$ (p-value) & $\widehat\beta$ (p-value) & $\widehat\beta$ (p-value)	& $\widehat\beta$ (p-value)	\cr
  	
\midrule

$\Delta=1$	&	0.037 (0.007)	&	-1.942 (0.037)	&	0.374 (0.507)	&	-0.252 (0.422)	&	0.030 (0.935)	&	-0.013 (0.874)	&	0.589 (0.027)	&	-0.291 (0.060)	&	0.105 (0.539)	&	-	&	-	\cr
$\Delta=2$	&	0.036 (0.004)	&	-2.281 (0.008)	&	0.316 (0.540)	&	-0.218 (0.459)	&	-0.031 (0.931)	&	0.004 (0.964)	&	0.580 (0.067)	&	-0.298 (0.048)	&	-	&	-0.062 (0.876)	&	-	\cr
$\Delta=3$	&	0.037 (0.008)	&	-2.213 (0.008)	&	0.434 (0.386)	&	-0.221 (0.433)	&	-0.047 (0.897)	&	0.009 (0.912)	&	0.566 (0.071)	&	-0.277 (0.064)	&	-	&	-0.069 (0.872)	&	-	\cr
$\Delta=4$	&	0.035 (0.009)	&	-2.034 (0.012)	&	0.663 (0.216)	&	-0.216 (0.433)	&	-0.057 (0.875)	&	0.013 (0.867)	&	0.521 (0.094)	&	-0.252 (0.083)	&	-	&	-0.016 (0.971)	&	-	\cr
$\Delta=5$	&	0.073 (0)	&	-2.365 (0.009)	&	0.276 (0.607)	&	-0.257 (0.379)	&	-0.071 (0.839)	&	0.015 (0.838)	&	0.493 (0.122)	&	-0.346 (0.040)	&	-	&	0.013 (0.977)	&	-	\cr
$\Delta=6$	&	0.069 (0.001)	&	-1.269 (0.148)	&	0.927 (0.142)	&	-0.381 (0.328)	&	0.201 (0.609)	&	-0.017 (0.836)	&	0.104 (0.739)	&	-	&	0.490 (0.104)	&	-0.426 (0.464)	&	-	\cr
$\Delta=7$	&	0.080 (0)	&	-2.087 (0.018)	&	0.421 (0.448)	&	-0.359 (0.275)	&	-0.147 (0.654)	&	0.025 (0.731)	&	0.631 (0.054)	&	-0.419 (0.028)	&	-	&	-0.026 (0.955)	&	-	\cr
$\Delta=8$	&	0.080 (0)	&	-2.090 (0.014)	&	0.410 (0.454)	&	-0.347 (0.286)	&	-0.141 (0.673)	&	0.019 (0.814)	&	0.715 (0.061)	&	-0.426 (0.027)	&	-	&	0.118 (0.837)	&	-	\cr
$\Delta=9$	&	0.097 (0)	&	-1.526 (0.076)	&	0.704 (0.224)	&	-0.467 (0.135)	&	0.379 (0.292)	&	-0.033 (0.692)	&	1.213 (0.051)	&	-	&	-	&	0.418 (0.583)	&	0.099 (0.036)	\cr
$\Delta=10$	&	0.097 (0)	&	-1.498 (0.078)	&	0.692 (0.232)	&	-0.461 (0.138)	&	0.378 (0.298)	&	-0.033 (0.695)	&	1.225 (0.048)	&	-	&	-	&	0.427 (0.572)	&	0.099 (0.035)	\cr
$\Delta=11$	&	0.108 (0)	&	-1.357 (0.091)	&	0.502 (0.376)	&	0.003 (0.994)	&	0.268 (0.468)	&	-0.004 (0.965)	&	0.988 (0.137)	&	-	&	-	&	0.408 (0.587)	&	0.088 (0.077)	\cr
$\Delta=12$	&	0.108 (0)	&	-1.350 (0.090)	&	0.478 (0.400)	&	0.005 (0.991)	&	0.269 (0.469)	&	-0.004 (0.961)	&	0.983 (0.143)	&	-	&	-	&	0.395 (0.603)	&	0.088 (0.079)	\cr
$\Delta=13$	&	0.111 (0)	&	-1.367 (0.088)	&	0.642 (0.314)	&	-0.122 (0.765)	&	0.294 (0.445)	&	-0.014 (0.878)	&	1.166 (0.095)	&	-	&	-	&	0.471 (0.524)	&	0.097 (0.064)	\cr
$\Delta=14$	&	0.121 (0)	&	-1.402 (0.112)	&	0.531 (0.421)	&	-0.099 (0.809)	&	0.290 (0.454)	&	-0.012 (0.896)	&	1.175 (0.098)	&	-	&	-	&	0.454 (0.550)	&	0.098 (0.066)	\cr

   \bottomrule
\end{tabular}
}
{\begin{flushleft} $-$ indicates the variable was not selected at $\Delta$ day.
\end{flushleft}}
\end{table}

\begin{table}
\small
\centering
\caption{
\bf  {Propensity Score Estimates of Reopening Business}}\label{tab:prop_business}
\resizebox{1.05\textwidth}{!}{
\begin{tabular}{lcccccccccc}
\toprule

 &
&\multicolumn{1}{c}{\bf Limit} &\multicolumn{1}{c}{\bf Multi-Unit} &\multicolumn{1}{c}{{\bf Mobile}} & &\multicolumn{1}{c}{\bf Per Capita}  & &\multicolumn{1}{c}{\bf No } &\multicolumn{1}{c}{\bf Male at Age 65} &
  \cr
  
  \multicolumn{1}{l}{{\bf Day}} &\multicolumn{1}{c}{{\boldmath $R_t$}} 
&\multicolumn{1}{c}{\bf English} &\multicolumn{1}{c}{\bf House} &\multicolumn{1}{c}{{\bf  Home}} &\multicolumn{1}{c}{\bf {Latino}} &\multicolumn{1}{c}{\bf Income}  &\multicolumn{1}{c}{\bf Disabled} &\multicolumn{1}{c}{\bf Vehicle} &\multicolumn{1}{c}{\bf and over} &\multicolumn{1}{c}{\bf Unemployed} 
  \cr

  	&	$\widehat\beta$ (p-value)		&	$\widehat\beta$ (p-value)		&	$\widehat\beta$ (p-value)		&	$\widehat\beta$ (p-value)		&	$\widehat\beta$ (p-value)		&	$\widehat\beta$$\times 10^{-4}$ (p-value)		&	$\widehat\beta$ (p-value) 	&	$\widehat\beta$ (p-value) 	&	$\widehat\beta$ (p-value) 	&	$\widehat\beta$ (p-value)	\cr
  	
\midrule
$\Delta=1$	&	-4.428 (0.001)	&	-0.257 (0.546)	&	-0.094 (0.645)	&	0.270 (0.039)	&	0.038 (0.583)	&	-0.036 (0.942)	&	-0.162 (0.265)	&	0.043 (0.769)	&	0.024 (0.900)	&	-	\cr
$\Delta=2$	&	-4.678 (0.001)	&	-0.290 (0.524)	&	-0.091 (0.674)	&	0.301 (0.022)	&	0.041 (0.561)	&	-0.044 (0.931)	&	-0.164 (0.254)	&	0.039 (0.791)	&	0.013 (0.945)	&	-	\cr
$\Delta=3$	&	-5.382 (0)	&	-0.322 (0.493)	&	-0.110 (0.627)	&	0.344 (0.005)	&	0.042 (0.555)	&	-0.024 (0.963)	&	-0.191 (0.172)	&	0.060 (0.688)	&	0.055 (0.771)	&	-	\cr
$\Delta=4$	&	-5.707 (0)	&	-0.329 (0.504)	&	-0.116 (0.632)	&	0.363 (0.004)	&	0.041 (0.567)	&	-0.028 (0.957)	&	-0.192 (0.175)	&	0.053 (0.727)	&	0.059 (0.754)	&	-	\cr
$\Delta=5$	&	-5.635 (0)	&	-0.342 (0.525)	&	-0.122 (0.649)	&	0.381 (0.003)	&	0.036 (0.633)	&	-0.010 (0.985)	&	-0.187 (0.221)	&	0.060 (0.723)	&	0.128 (0.491)	&	-	\cr
$\Delta=6$	&	-6.222 (0)	&	-0.409 (0.470)	&	-0.123 (0.663)	&	0.432 (0.003)	&	0.042 (0.577)	&	0.055 (0.923)	&	-0.206 (0.191)	&	0.071 (0.692)	&	0.146 (0.440)	&	-	\cr
$\Delta=7$	&	-6.572 (0)	&	-0.475 (0.464)	&	-0.126 (0.693)	&	0.427 (0.005)	&	0.054 (0.524)	&	-0.073 (0.899)	&	-0.222 (0.168)	&	0.070 (0.707)	&	0.133 (0.496)	&	-	\cr
$\Delta=8$	&	-8.161 (0)	&	-0.362 (0.641)	&	-0.325 (0.419)	&	0.612 (0)	&	0.065 (0.492)	&	0.514 (0.416)	&	-0.198 (0.218)	&	0.193 (0.337)	&	0.167 (0.471)	&	-0.641 (0.111)	\cr
$\Delta=9$	&	-8.768 (0)	&	-0.443 (0.602)	&	-0.308 (0.473)	&	0.645 (0)	&	0.073 (0.479)	&	0.345 (0.478)	&	-0.207 (0.259)	&	0.181 (0.384)	&	0.119 (0.630)	&	-0.624 (0.131)	\cr
$\Delta=10$	&	-8.853 (0)	&	-0.440 (0.613)	&	-0.314 (0.468)	&	0.679 (0)	&	0.070 (0.502)	&	0.345 (0.600)	&	-0.251 (0.165)	&	0.177 (0.402)	&	0.099 (0.697)	&	-0.640 (0.131)	\cr
$\Delta=11$	&	-9.031 (0)	&	-0.456 (0.598)	&	-0.290 (0.502)	&	0.665 (0)	&	0.069 (0.501)	&	0.224 (0.735)	&	-0.253 (0.161)	&	0.145 (0.504)	&	0.080 (0.753)	&	-0.670 (0.118)	\cr
$\Delta=12$	&	-8.949 (0)	&	-0.485 (0.578)	&	-0.271 (0.525)	&	0.676 (0)	&	0.071 (0.488)	&	0.174 (0.793)	&	-0.285 (0.122)	&	0.128 (0.552)	&	0.058 (0.819)	&	-0.656 (0.131)	\cr
$\Delta=13$	&	-9.090 (0)	&	-0.496 (0.570)	&	-0.263 (0.541)	&	0.668 (0)	&	0.072 (0.484)	&	0.141 (0.838)	&	-0.304 (0.133)	&	0.108 (0.635)	&	0.060 (0.822)	&	-0.700 (0.136)	\cr
$\Delta=14$	&	-9.278 (0)	&	-0.497 (0.560)	&	-0.255 (0.539)	&	0.668 (0)	&	0.068 (0.504)	&	0.070 (0.920)	&	-0.323 (0.115)	&	0.088 (0.700)	&	0.044 (0.871)	&	-0.688 (0.146)	\cr

   \bottomrule
\end{tabular}
}
{\begin{flushleft} $-$ indicates the variable was not selected at $\Delta$ day.
\end{flushleft}}
\end{table}

\begin{table}
\small
\centering
\caption{
\bf  {Propensity Score Estimates of Reopening Restaurants}}\label{tab:prop_restaurant}
\resizebox{1.05\textwidth}{!}{
\begin{tabular}{lcccccccccc}
\toprule

  & &\multicolumn{1}{c}{{\bf Per Capita}}  &\multicolumn{1}{c}{\bf Multi-Unit}   &\multicolumn{1}{c}{\bf Limit} &\multicolumn{1}{c}{\bf Mobile} &  &\multicolumn{1}{c}{\bf Single } &\multicolumn{1}{c}{\bf No} &
  \cr

 \multicolumn{1}{l}{{\bf Day}} &\multicolumn{1}{c}{{\bf {$R_t$}}} 
 &\multicolumn{1}{c}{{\bf Income}}  &\multicolumn{1}{c}{\bf House}  &\multicolumn{1}{c}{\bf English} &\multicolumn{1}{c}{\bf Home}  &\multicolumn{1}{c}{\bf Latino} &\multicolumn{1}{c}{\bf Parent House} &\multicolumn{1}{c}{\bf  Vehicle}  &\multicolumn{1}{c}{\bf Disabled} &\multicolumn{1}{c}{\bf Poverty}
  \cr

  	&	$\widehat\beta$ (p-value)		&	$\widehat\beta$$\times 10^{-4}$ (p-value)		&	$\widehat\beta$ (p-value)		&	$\widehat\beta$ (p-value)		&	$\widehat\beta$ (p-value)		&	$\widehat\beta$ (p-value)		&	$\widehat\beta$ (p-value) 	&	$\widehat\beta$ (p-value) 	&	$\widehat\beta$ (p-value) 	&	$\widehat\beta$ (p-value) 	\cr
  	
\midrule

$\Delta=1$	&	-2.094 (0.039)	&	-0.362 (0.681)	&	-0.054 (0.782)	&	-0.140 (0.613)	&	0.085 (0.495)	&	0.004 (0.923)	&	0.113 (0.886)	&	-0.034 (0.850)	&	-0.227 (0.272)	&	0.114 (0.573)	\cr
$\Delta=2$	&	-2.176 (0.035)	&	-0.379 (0.664)	&	-0.053 (0.784)	&	-0.159 (0.568)	&	0.071 (0.567)	&	0.007 (0.875)	&	0.108 (0.891)	&	-0.045 (0.805)	&	-0.226 (0.281)	&	0.126 (0.528)	\cr
$\Delta=3$	&	-2.190 (0.036)	&	-0.444 (0.616)	&	-0.048 (0.811)	&	-0.137 (0.629)	&	0.089 (0.470)	&	0.000 (0.998)	&	0.163 (0.839)	&	-0.053 (0.772)	&	-0.250 (0.250)	&	0.126 (0.535)	\cr
$\Delta=4$	&	-2.251 (0.034)	&	-0.441 (0.621)	&	-0.050 (0.799)	&	-0.119 (0.676)	&	0.101 (0.395)	&	-0.004 (0.925)	&	0.208 (0.798)	&	-0.065 (0.726)	&	-0.256 (0.251)	&	0.138 (0.510)	\cr
$\Delta=5$	&	-2.624 (0.033)	&	-0.451 (0.613)	&	-0.070 (0.716)	&	-0.135 (0.626)	&	0.124 (0.298)	&	-0.005 (0.905)	&	0.148 (0.864)	&	-0.063 (0.741)	&	-0.315 (0.174)	&	0.159 (0.463)	\cr
$\Delta=6$	&	-2.740 (0.030)	&	-0.482 (0.597)	&	-0.057 (0.768)	&	-0.145 (0.606)	&	0.130 (0.283)	&	-0.005 (0.921)	&	0.231 (0.791)	&	-0.079 (0.686)	&	-0.318 (0.175)	&	0.159 (0.475)	\cr
$\Delta=7$	&	-2.823 (0.028)	&	-0.539 (0.558)	&	-0.068 (0.727)	&	-0.154 (0.576)	&	0.134 (0.285)	&	-0.001 (0.974)	&	0.282 (0.753)	&	-0.074 (0.708)	&	-0.320 (0.158)	&	0.145 (0.512)	\cr
$\Delta=8$	&	-2.873 (0.027)	&	-0.679 (0.473)	&	-0.066 (0.745)	&	-0.155 (0.582)	&	0.138 (0.266)	&	-0.002 (0.963)	&	0.331 (0.728)	&	-0.073 (0.714)	&	-0.333 (0.145)	&	0.133 (0.555)	\cr
$\Delta=9$	&	-3.035 (0.029)	&	-0.781 (0.431)	&	-0.076 (0.720)	&	-0.163 (0.565)	&	0.143 (0.242)	&	-0.005 (0.908)	&	0.451 (0.661)	&	-0.071 (0.730)	&	-0.374 (0.097)	&	0.143 (0.532)	\cr
$\Delta=10$	&	-3.565 (0.037)	&	-0.950 (0.352)	&	-0.083 (0.701)	&	-0.177 (0.542)	&	0.153 (0.188)	&	-0.003 (0.938)	&	0.504 (0.624)	&	-0.067 (0.742)	&	-0.410 (0.080)	&	0.128 (0.593)	\cr
$\Delta=11$	&	-3.596 (0.034)	&	-0.965 (0.342)	&	-0.081 (0.711)	&	-0.183 (0.536)	&	0.153 (0.203)	&	-0.004 (0.920)	&	0.545 (0.600)	&	-0.071 (0.732)	&	-0.407 (0.075)	&	0.124 (0.600)	\cr
$\Delta=12$	&	-3.786 (0.029)	&	-1.018 (0.330)	&	-0.077 (0.730)	&	-0.195 (0.516)	&	0.140 (0.250)	&	-0.004 (0.926)	&	0.602 (0.569)	&	-0.085 (0.687)	&	-0.406 (0.082)	&	0.125 (0.595)	\cr
$\Delta=13$	&	-3.860 (0.027)	&	-1.018 (0.314)	&	-0.086 (0.713)	&	-0.190 (0.524)	&	0.131 (0.285)	&	-0.005 (0.901)	&	0.537 (0.640)	&	-0.091 (0.677)	&	-0.408 (0.105)	&	0.136 (0.572)	\cr
$\Delta=14$	&	-4.028 (0.024)	&	-1.180 (0.287)	&	-0.099 (0.669)	&	-0.177 (0.563)	&	0.118 (0.337)	&	-0.010 (0.819)	&	0.700 (0.562)	&	-0.097 (0.659)	&	-0.432 (0.093)	&	0.140 (0.572)	\cr

   \bottomrule
\end{tabular}
}
\end{table}

\begin{table}
\small
\centering
\caption{
\bf  {Propensity Score Estimates of Reopening Bars}}\label{tab:prop_bar}
\resizebox{1.05\textwidth}{!}{
\begin{tabular}{lccccccccc}
\toprule

 &\multicolumn{1}{c}{{\bf New}} 
 &\multicolumn{1}{c}{\bf New}  &\multicolumn{1}{c}{\bf Limit} &\multicolumn{1}{c}{\bf Multi-Unit} &\multicolumn{1}{c}{\bf Per Capita}  & &\multicolumn{1}{c}{\bf Mobile} &
  \cr

\multicolumn{1}{l}{{\bf Day}} &\multicolumn{1}{c}{{\bf Case}} 
 &\multicolumn{1}{c}{\bf Death}  &\multicolumn{1}{c}{\bf English} &\multicolumn{1}{c}{\bf  House}  &\multicolumn{1}{c}{\bf Income}  &\multicolumn{1}{c}{\bf Latino} &\multicolumn{1}{c}{\bf Home} &\multicolumn{1}{c}{\bf Institutionalized} 
  \cr

  	&	$\widehat\beta$ (p-value)		&	$\widehat\beta$ (p-value)		&	$\widehat\beta$ (p-value)		&	$\widehat\beta$ (p-value)		&	$\widehat\beta\times 10^{-4}$ (p-value)		&	$\widehat\beta$ (p-value)		&	$\widehat\beta$ (p-value) 	&	$\widehat\beta$ (p-value) 		\cr
  	
\midrule
$\Delta=1$	&	-0.060 (0.216)	&	-0.631 (0.378)	&	-0.320 (0.517)	&	0.032 (0.922)	&	-0.330 (0.648)	&	0.032 (0.560)	&	-0.066 (0.708)	&	0.270 (0.530)	\cr
$\Delta=2$	&	-0.062 (0.197)	&	-0.598 (0.398)	&	-0.339 (0.498)	&	0.033 (0.920)	&	-0.373 (0.607)	&	0.032 (0.549)	&	-0.076 (0.673)	&	0.270 (0.523)	\cr
$\Delta=3$	&	-0.063 (0.194)	&	-0.602 (0.394)	&	-0.345 (0.486)	&	0.038 (0.907)	&	-0.442 (0.549)	&	0.033 (0.535)	&	-0.078 (0.667)	&	0.258 (0.541)	\cr
$\Delta=4$	&	-0.061 (0.209)	&	-0.626 (0.372)	&	-0.321 (0.528)	&	0.028 (0.934)	&	-0.412 (0.582)	&	0.030 (0.583)	&	-0.072 (0.692)	&	0.254 (0.544)	\cr
$\Delta=5$	&	-0.062 (0.198)	&	-0.629 (0.370)	&	-0.319 (0.542)	&	0.022 (0.948)	&	-0.423 (0.575)	&	0.028 (0.605)	&	-0.070 (0.701)	&	0.231 (0.578)	\cr
$\Delta=6$	&	-0.065 (0.179)	&	-0.654 (0.346)	&	-0.307 (0.558)	&	0.007 (0.984)	&	-0.377 (0.617)	&	0.029 (0.601)	&	-0.073 (0.700)	&	0.298 (0.482)	\cr
$\Delta=7$	&	-0.068 (0.150)	&	-0.632 (0.359)	&	-0.338 (0.532)	&	0.005 (0.990)	&	-0.367 (0.632)	&	0.033 (0.553)	&	-0.085 (0.653)	&	0.283 (0.499)	\cr
$\Delta=8$	&	-0.069 (0.148)	&	-0.644 (0.353)	&	-0.365 (0.500)	&	0.012 (0.973)	&	-0.361 (0.640)	&	0.035 (0.524)	&	-0.081 (0.674)	&	0.294 (0.486)	\cr
$\Delta=9$	&	-0.071 (0.149)	&	-0.674 (0.359)	&	-0.468 (0.403)	&	0.056 (0.873)	&	-0.437 (0.589)	&	0.044 (0.427)	&	-0.073 (0.715)	&	0.301 (0.500)	\cr
$\Delta=10$	&	-0.072 (0.138)	&	-0.658 (0.371)	&	-0.475 (0.406)	&	0.058 (0.870)	&	-0.475 (0.559)	&	0.044 (0.423)	&	-0.075 (0.711)	&	0.305 (0.485)	\cr
$\Delta=11$	&	-0.077 (0.117)	&	-0.662 (0.365)	&	-0.489 (0.399)	&	0.044 (0.901)	&	-0.495 (0.543)	&	0.045 (0.420)	&	-0.091 (0.660)	&	0.340 (0.455)	\cr
$\Delta=12$	&	-0.076 (0.117)	&	-0.711 (0.331)	&	-0.495 (0.407)	&	0.037 (0.920)	&	-0.544 (0.515)	&	0.044 (0.444)	&	-0.093 (0.658)	&	0.313 (0.491)	\cr
$\Delta=13$	&	-0.077 (0.113)	&	-0.718 (0.321)	&	-0.517 (0.379)	&	0.045 (0.900)	&	-0.562 (0.500)	&	0.047 (0.410)	&	-0.098 (0.641)	&	0.337 (0.447)	\cr
$\Delta=14$	&	-0.079 (0.110)	&	-0.760 (0.311)	&	-0.567 (0.346)	&	0.049 (0.892)	&	-0.645 (0.447)	&	0.051 (0.376)	&	-0.119 (0.589)	&	0.284 (0.528)	\cr

   \bottomrule
\end{tabular}
}
\end{table}

\subsection*{Web Appendix B: HTE Results}
\begin{figure}[!ht]
\centering
\includegraphics[width=0.6\textwidth]{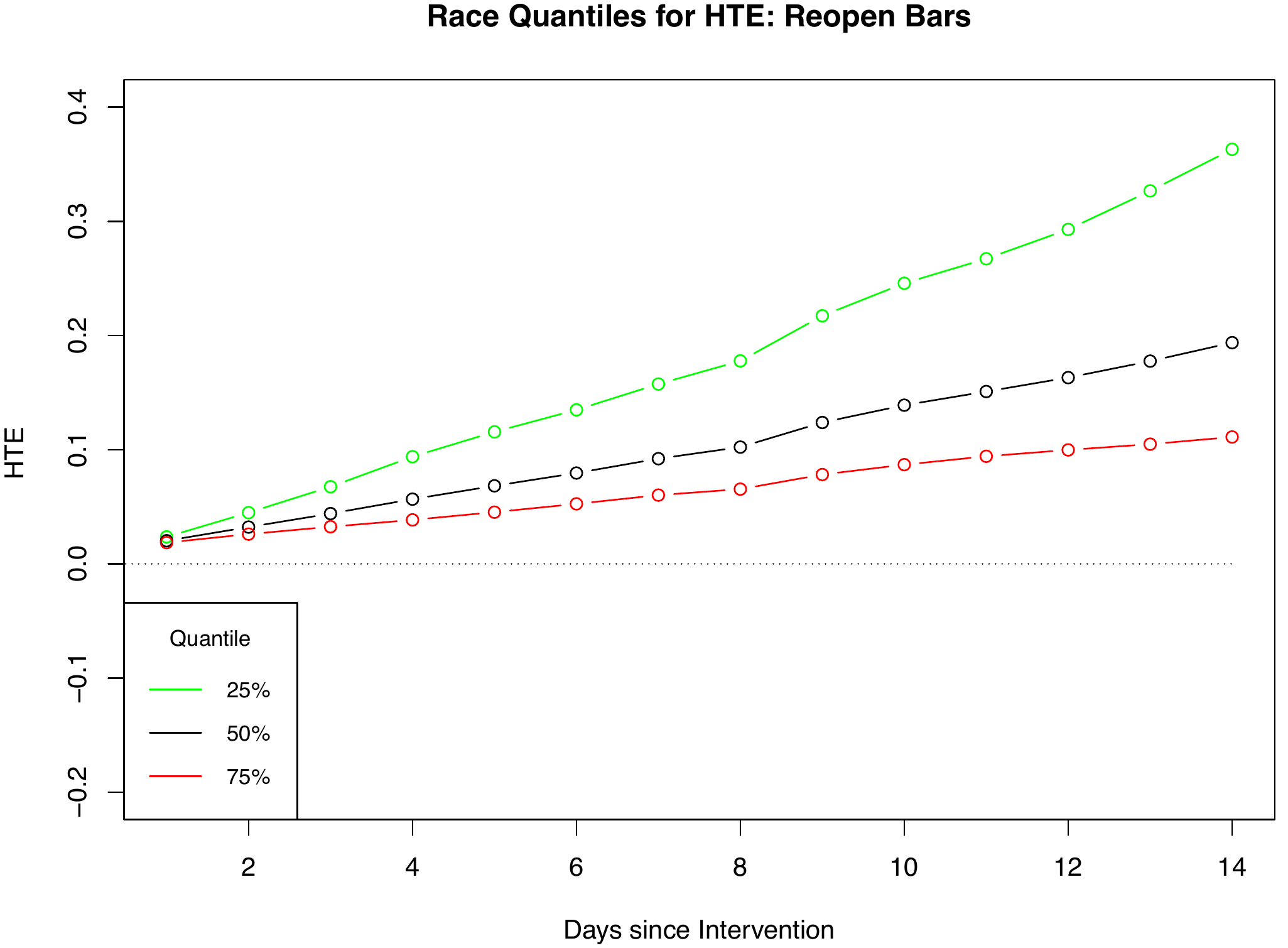}
\caption{HTE of White for the NPI: Reopening bars. The effects are estimated with the moderator fixed at a given quantile  and other variables fixed at the mean level.}\label{fig:HTE}

\end{figure}

This section shows the estimated  HTE of race (percentage of White) for reopening bars (Figure ~\ref{fig:HTE}). The effects are estimated with moderator fixed at a given quantile (e.g., 25th percentile, 50th percentile, 75th percentile) over all states and other covariates fixed at the  mean level. 

\clearpage

\end{document}